\documentclass[10pt,journal,compsoc]{IEEEtran}

\ifCLASSOPTIONcompsoc
  \usepackage[nocompress]{cite}
\else
  \usepackage{cite}
\fi
\ifCLASSINFOpdf
\else
\fi

\usepackage{array}
\usepackage{makecell}
\usepackage{multirow}
\usepackage{color}
\usepackage{diagbox}
\usepackage{tikz}
\usepackage{xspace}
\usepackage{subfigure} 
\usepackage{pgf-pie}
\usepackage{graphicx}
\usepackage{subfigure}
\usepackage{fancyhdr}
\usetikzlibrary{shapes.geometric,calc}
\newcommand\score[2]{
\pgfmathsetmacro\pgfxa{#1+1}
\tikzstyle{scorestars}=[star, star points=5, star point ratio=2.25, draw,inner sep=0.15em,anchor=outer point 3]
\begin{tikzpicture}[baseline]
  \foreach \i in {1,...,#2} {
    \pgfmathparse{(\i<=#1?"black":"white")}
    \edef\starcolor{\pgfmathresult}
    \draw (\i*1em,0) node[name=star\i,scorestars,fill=\starcolor]  {};
   }
   \pgfmathparse{(#1>int(#1)?int(#1+1):0}
   \let\partstar=\pgfmathresult
   \ifnum\partstar>0
     \pgfmathsetmacro\starpart{#1-(int(#1))}
     \path [clip] ($(star\partstar.outer point 3)!(star\partstar.outer point 2)!(star\partstar.outer point 4)$) rectangle 
    ($(star\partstar.outer point 2 |- star\partstar.outer point 1)!\starpart!(star\partstar.outer point 1 -| star\partstar.outer point 5)$);
     \fill (\partstar*1em,0) node[scorestars,fill=black]  {};
   \fi

,\end{tikzpicture}
}

\newcommand{\realize}{fulfill\xspace}
\newcommand{\realizing}{fulfilling\xspace}

\newcommand{\realized}{fulfilled\xspace}
\newcommand{\realization}{fulfillment\xspace}

\begin{document}

\title{No Free Lunch: Microservice Practices Reconsidered in Industry}


\author{Qilin~Xiang,
    Xin~Peng,
    Chuan~He,
    Hanzhang~Wang,
        Tao~Xie,
        Dewei~Liu,
        Gang~Zhang,
        and~Yuanfang~Cai
        
 \IEEEcompsocitemizethanks{
\IEEEcompsocthanksitem X. Peng is the corresponding author.
\IEEEcompsocthanksitem Q. Xiang, X. Peng, C. He, and D. Liu are with the School of Computer Science and the Shanghai Key Laboratory of Data Science, Fudan University, Shanghai, China, and Shanghai Institute of Intelligent Electronics \& Systems, China.
\IEEEcompsocthanksitem H. Wang is with the eBay Inc., USA.
\IEEEcompsocthanksitem T. Xie is with the Peking University, China.
\IEEEcompsocthanksitem G. Zhang is with the Emergent Design Inc., China.
\IEEEcompsocthanksitem Y. Cai is with the Drexel University, USA.
\protect\\
}
\thanks{}}

\markboth{IEEE TRANSACTIONS ON SOFTWARE ENGINEERING,~Vol.~XX, No.~XX, XX~2020}
{Shell \MakeLowercase{\textit{et al.}}: Bare Demo of IEEEtran.cls for Computer Society Journals}

\IEEEtitleabstractindextext{
\begin{abstract}
Microservice architecture advocates a number of technologies and practices such as lightweight container, container orchestration, and DevOps, with the promised benefits of faster delivery, improved scalability, and greater autonomy.
However, microservice systems implemented in industry vary a lot in terms of adopted practices and achieved benefits, drastically different from what is advocated in the literature.
In this article, we conduct an empirical study, including an online survey with 51 responses and 14 interviews for experienced microservice experts to advance our understanding regarding to microservice practices in industry.
As a part of our findings, the empirical study clearly revealed three levels of maturity of  microservice systems (from basic to advanced): independent development and deployment, high scalability and availability, and service ecosystem, categorized by the \realized benefits of microservices. 
We also identify 11 practical issues that constrain the microservice capabilities of organizations. 
For each issue, we summarize the practices that have been explored and adopted in industry, along with the remaining challenges.
Our study can help practitioners better position their microservice systems and determine what infrastructures and capabilities are worth investing.
Our study can also help researchers better understand industrial microservice practices and identify useful research problems.
\end{abstract}

}
\maketitle
\IEEEdisplaynontitleabstractindextext
\IEEEpeerreviewmaketitle

\IEEEraisesectionheading{\section{Introduction}\label{sec:introduction}}



\thispagestyle{plain}

\IEEEPARstart{M}{icroservice} architecture is an architectural style that structures an  application as a suite of loosely coupled services, each of which has a single responsibility and can be deployed independently, scaled independently, and tested independently~\cite{MICROSERVICE, IEEESoft15MS}.
Different from traditional service-oriented architecture (SOA),  which is viewed mostly as an integration solution, microservices are individual software applications that communicate with each other through well-defined network interfaces~\cite{S18JSFCA}.
Microservice architecture is supposed to deliver the benefits of faster delivery, improved scalability, and greater autonomy~\cite{MICROSERVICE}, and has been the latest trend in building cloud native applications.
Many Internet applications (e.g.,  Amazon~\cite{AMAZON}, Netflix~\cite{NETFLIX}, Tencent's WeChat~\cite{WECHAT}, eBay's developers program~\cite{ebay}) and enterprise applications (e.g., online meeting applications and BPM applications~\cite{TSE18MS}) have been built based on microservice architecture.

Microservice architecture advocates a series of technologies and practices. An individual microservice is usually packaged and deployed in the cloud using a lightweight container (e.g., Docker~\cite{DOCKER}). A collection of microservices are typically managed using container orchestration technologies (e.g., Kubernetes~\cite{K8S}, Docker Swarm~\cite{DockerSwarm}, and Mesos~\cite{MESOS}), following industry-proven DevOps practices and supported by fully automated software integration and delivery machinery~\cite{S18JSFCA, IEEE15CPC, BASS15DEVOPS, HUMBLE10CONTINOUS}.
Recently there is a trend of applying more advanced microservice technologies and practices such as chaos engineering~\cite{ChaosEngineering}, serverless~\cite{SERVERLESS}, and service mesh~\cite{ServiceMesh}.

More and more organizations, with various sizes and domains, are adopting microservice architecture. 
These organizations have exhibited drastic differences in terms of their adopted technologies and practices, as well as achieved benefits. 
The adoption of all recommended technologies and practices would require great investments in process and infrastructure, and the adoption of new technologies and practices involves uncertainties and risks.
Therefore, the organizations usually selectively adopt part of the recommanded microservice technologies and practices.

On the other hand, the organizations' motivation to pursue more sophisticated infrastructure is usually driven by business needs from market and internal forces.
For example, an organization running a system with stable and predictable access load may not be interested in flexible scalability, but could be interested in fast delivery of new features. Therefore, an organization will only make more investment on advanced microservice infrastructure when the corresponding driving forces are taking effects.

Considering the sharp contrast between the advocated benefits and practices in literature and the \realized benefits and adopted practices in industry, we are curious to explore microservice practices in industry to find out how and why they differ.
Understanding the differences among the maturity levels of microservice systems in practice is important for practitioners to position their systems and to create a roadmap of continuous improvement, and for researchers to understand the difficulties and challenges at different levels.

Some previous research efforts~\cite{SOCA16SMSMA, ICSA17RAMTFPIA, CCSS16MSMS} conduct systematic mapping studies to learn the characteristics, benefits, and research trends of microservices from the literature, but not directly reflecting the microservice practices in industry. 
Other research efforts~\cite{ICSA18MTMAAIS, ICC17PMIMMAAEI, ICSA19MARII} report industrial surveys that reveal common practices, benefits, and challenges.
Some empirical studies~\cite{Taibi2020, IEEESoftware18ODMBS, msa-antipattern-pitfalls} further identify microservice-specific bad smells and anti-patterns from both organizational and technical perspectives.
These previous studies do not reveal the existence of maturity levels and their associated practices and benefits. 
As a result, the practitioners cannot position their current practice, or make informed decisions on if and what improvement is needed to achieve higher level of maturity. 

In this article, we report our empirical study to investigate industrial microservice systems with a special focus on the differences in terms of adopted practices and \realized benefits.
Based on a conceptual model to characterize microservice maturity levels, we investigate the following three research questions.

\begin{itemize}
\item \textbf{RQ1 (Maturity Level)}:
What maturity levels can industrial microservice systems be classified into?
What are the characteristics and driving forces of each level?
What benefits and promises can be achieved at each level?

\item \textbf{RQ2 (Issue)}:
What are the common issues that the microservice development capabilities are subject to?
How do these common issues influence the microservice systems at different maturity levels?

\item \textbf{RQ3 (Practice and Challenge)}:
What practices have been explored and adopted for the identified issues at each maturity level?
How well are the issues addressed by these practices and what challenges still remain there?

\end{itemize}

To answer the these questions, we first conduct an online survey to gain an overview of industrial microservice systems, serving as the basis for the next phase interviews.
The data collected from the survey can be found in our replication package\footnote{https://replication-package-tse.github.io/TSE2020/index.html}.
We then conduct a series of interviews with the architects and technical leaders for some of these systems to collect high-fidelity and in-depth data around microservice practices and benefits.

Based on the survey and interviews, we observed that the practices and benefits that have been adopted and \realized vary greatly in different systems.
Some systems have just \realized the basic benefits of independent development and deployment, but may still well support the business needs of the organizations.
In contrast, some other systems have been evolved into service ecosystems and achieved high scalability, availability, and expandability of business domains, driven by the forces of business expansion and merger. 
Most interestingly, these differences can be clearly characterized by three levels of  maturity, from basic to advanced: (1) independent development and deployment, (2) high scalability and availability, and (3) service ecosystem.

The success of microservice architecture relies on a collection of capabilities such as service decomposition, logging and monitoring, fault localization, and service evolution.
These capabilities are often restricted by various issues raised in microservice development, and thus influence the \realization of microservice benefits.
Based on the survey and interviews, we identified 11 common issues and analyzed their influences on microservice systems at different levels.
For each issue, we summarize the practices that have been explored and adopted, and the remaining challenges.

Our findings in this study are valuable for both practitioners and researchers.
For practitioners, our findings can help them  position their microservice systems at a proper level according to the business needs and determine what infrastructures and capabilities are worth investing.
More concretely, they can learn the issues that restrict their capabilities and the practices that they can follow to address the issues.
For researchers, our findings can help them understand the situation of industrial microservice practices and the needs of microservice systems of different levels.
They may identify valuable research to address the challenges in the current microservice practices and have a better understanding of problems in the context of microservices.

The rest of the article is organized as follows. 
Section~\ref{sec:studydesign} introduces the conceptual model and the study process. 
Section~\ref{sec:results} reports the data statistics of survery results. 
Section~\ref{sec:findings} summarizes our findings and answers the three research questions based on the survey and interviews. 
Section~\ref{sec:discuss} discusses the research opportunities at different maturity levels.
Section~\ref{sec:threat} and~\ref{sec:related} describe threats to validity and related work, respectively.  
Section~\ref{sec:conclude} concludes this article with the future work.

\section{Study Design}\label{sec:studydesign}
We design the research questions and study process  based on a collection of concepts related to the maturity levels.
In this section, we first introduce the conceptual model and then the study process.

\subsection{Conceptual Model}\label{sec:concept}
\begin{figure}
    \centering
    \includegraphics[width=1.0\columnwidth]{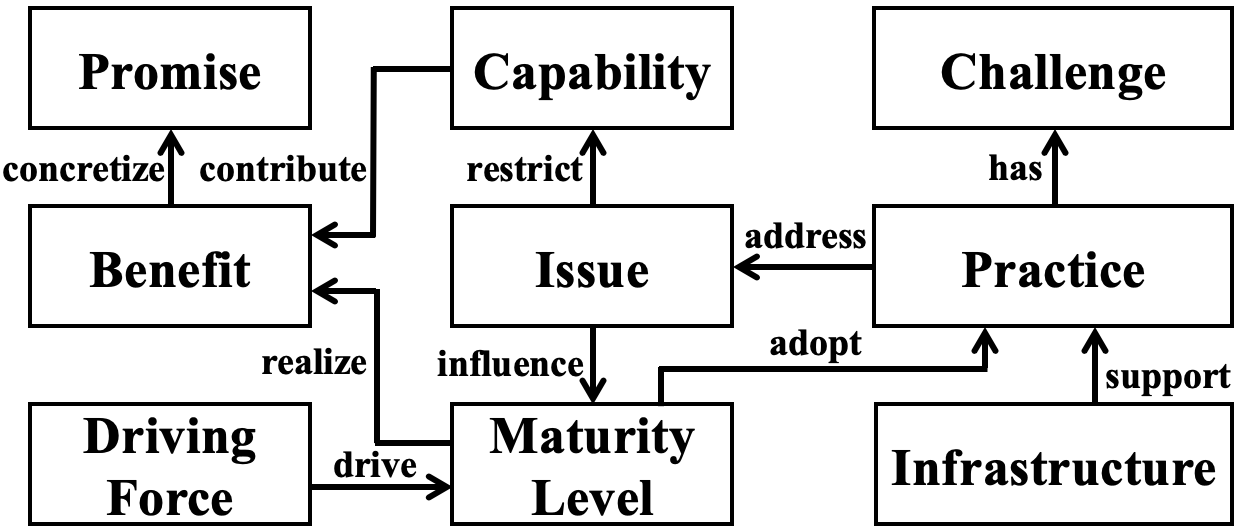}
    \caption{Conceptual Model} \label{fig:concept}
\end{figure}

Our conceptual model is shown in Figure~\ref{fig:concept}.
Our study starts with the following three important promises that are often expected from a microservice architecture~\cite{S18JSFCA}.
\begin{itemize}
\item \textbf{Faster Delivery}. Ideas can be turned into features running in production in a shorter time.

\item \textbf{Improved Scalability and Availability}. The system can better scale with environmental changes (e.g., system load) and at the same time ensure the availability.

\item \textbf{Greater Autonomy}. Development teams can make technical decisions in a more autonomous way.
\end{itemize}

These promises can be concretized into a set of benefits~\cite{MICROSERVICE, JSS18PGMSGLR, S18JSFCA, CCSS16MSMS, SOCA16SMSMA} as shown in Table~\ref{tab:benefit}.
The \realization of these benefits varies greatly in industrial microservice systems.
These systems can be classified into different maturity levels based on their achieved benefits.
Note that the \realization of a benefit implies the \realization of the business value behind, which is different from the satisfaction on the corresponding aspect.
For example, an organization does not \realize the benefit of flexible and automatic scalability in a microservice system, but may still be satisfied with its scalability if the number of users accessing the system remains stable. 
Investment in higher maturity level of microservice is often driven by some external forces, which usually originates from the requests of business development, e.g., the increase of service requests and the expansion of business domain.
An organization may choose to make a microservice system stay at a lower level if there are no business requests driving them to change.

\begin{table}[t]
  \scriptsize
  \centering
  \caption{Benefits of Microservices}\label{tab:benefit}
 \begin{tabular}{|c|c|m{4.0cm}|}
  \hline
    \multicolumn{1}{|c|}{\textbf{Promise}} &  \multicolumn{1}{c|}{\textbf{Benefit}} &  \multicolumn{1}{c|}{\textbf{Description}} \\
    \hline
    \multirow{2}{*}{\shortstack{Faster\\Delivery}} & \shortstack{Parallel\\Development} & Different services can be developed and deployed independently and in parallel.  \\
    \cline{2-3} & \shortstack{Extendibility\\and\\Expandability} & The system can well support the extension of new requirements and the expansion of business areas.\\
    \hline
    \multirow{2}{*}{\shortstack{Improved\\Scalability\\and\\Availability} }& \shortstack{Flexible\\and\\Automatic Scalability} & The system can automatically and flexibly scale services according to the changes of load and environment. \\
    \cline{2-3} & \shortstack{Fault Tolerance\\and\\Fault Isolation}& The system can ensure high availability by fault tolerance and fault isolation.  \\
    \hline
    \multirow{2}{*}{\shortstack{Greater\\Autonomy}} & \shortstack{Reduced\\Communication Cost} & 
Different teams can make technical decisions independently, and thus require less communication between teams.  \\
    \cline{2-3} & \shortstack{Flexible Choice\\of\\Technology Stack} & Different teams can flexibly choose the most appropriate technology stack (e.g., programming languages, frameworks) for their services.\\
     \hline
  \end{tabular}
\end{table}

The \realization of the benefits highly relies on a collection of capabilities of the organizations~\cite{MICROSERVICE, MSPATTERN, ICC17PMIMMAAEI, ICSA18MTMAAIS, ICSA19MARII}, including service decomposition, database decomposition, deployment, service communication design, API gateway design, service registration and discovery, logging and monitoring, performance and availability assurance, testing, fault localization, and service evolution.
These capabilities are restricted by various issues that are specific to microservice systems.
An issue has different impact on systems with  different maturity levels.
Accordingly, for microservice systems at different maturity levels, the developers have explored and adopted some practices to address the issues.
Some practices are associated with specific advances on the microservice infrastructures.
These practices alleviate the related issues and can be followed by other microservice systems, but may still have challenges to be further addressed.
For example, after a strategy of database decomposition is used to cope with database performance issues, there may still exist data consistency challenges.

\subsection{Study Process}
\begin{figure*}[!h]
    \centering
    \includegraphics[width=2.1\columnwidth]{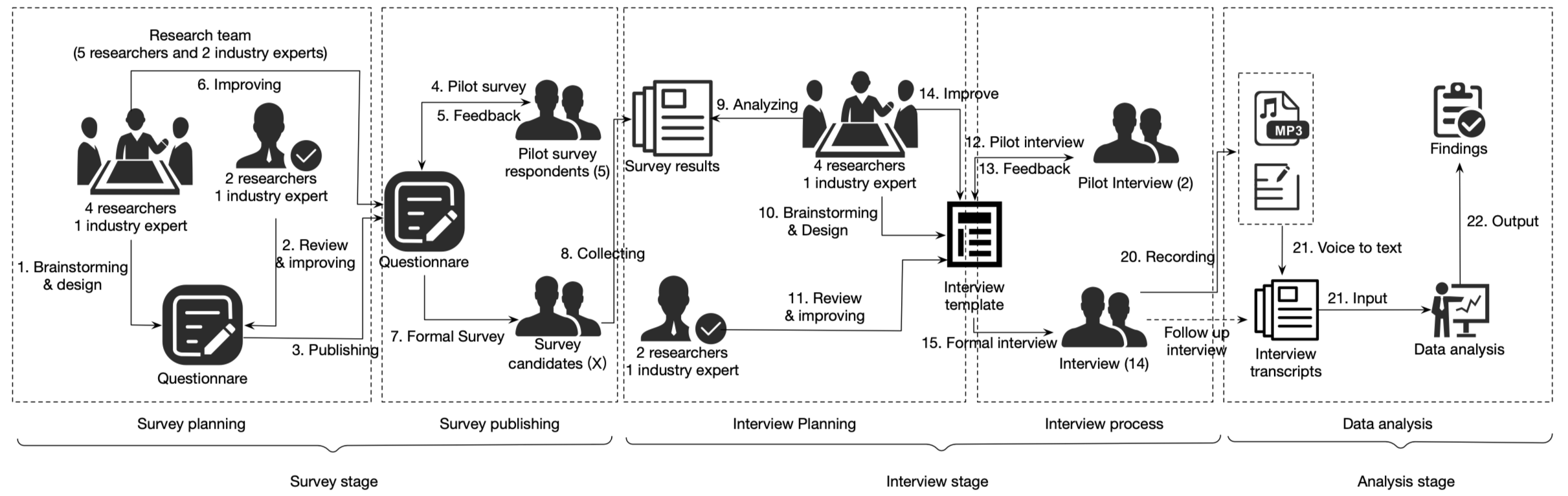}
    \caption{Study Process}
    \label{fig:study-process}
\end{figure*}

\begin{figure}[!h]
    \centering
    \includegraphics[width=1.0\columnwidth]{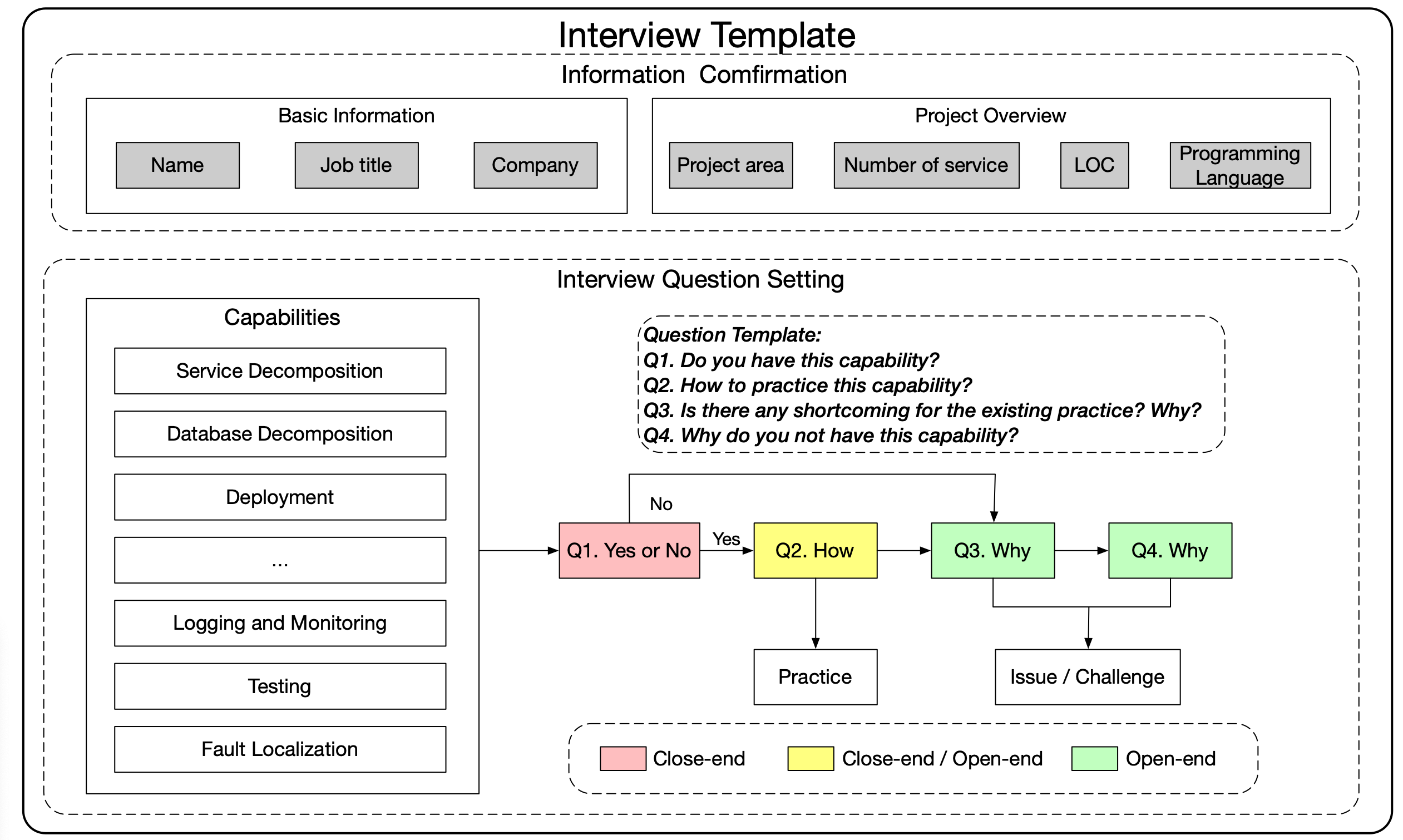}
    \caption{Interview Template}
    \label{fig:interview-template}
\end{figure}

Figure~\ref{fig:study-process} illustrates the three main stages of this study: survey, interview, and analysis. 
Our team is composed of six researchers (including three professors) and two industry experts. Each of the three main stages of this study has several key objectives:

\begin{itemize}
\item \textbf{Survey}: 
Collect objective and quantifiable data to gain an overview of multiple microservice projects and help identify the interviewee candidates. In addition, the survey provides a basis for the interview template shown in Figure~\ref{fig:interview-template}. 

\item \textbf{Interview}:
Collect high-fidelity and in-depth data around microservice practices of the IT industries. In this stage, we focus on the \realized benefits, capabilities and related issues, infrastructures and other practices, as well as remaining challenges of the subject systems.

\item \textbf{Analysis}:
Through descriptive statistics and quantitative analysis, identify practice patterns and issue trends from data collected from both the survey and interviews. In this stage, we finalize the identified maturity levels (RQ1) and learn how issues, practices, and challenges vary at different levels (RQ2, RQ3).
\end{itemize}

\subsubsection{Survey}
The survey stage collects technical details to create an overall picture of microservice capabilities.

As shown in Figure~\ref{fig:study-process}, at the survey stage, four researchers and one industry expert conducted brainstorming to initially design the survey questionnaire based on the required capabilities of microservices, then two researchers and one industry expert review and improve the questionnare.
To evaluate and improve the survey questionnaire, we rigorously conducted five pilot surveys with experienced industry experts, resulting in five revisions based on the feedback.

We then published the questionnaire\footnote{https://forms.gle/bHjnGmQnB2ddcG8E9} on social media (e.g., Twitter and WeChat) and microservice-related technology communities (e.g., the DevOps community and Kubernetes community). We also sent invitations to a manually-prepared mailing list from microservice related published articles in technical forums (e.g., InfoQ) and microservice related opensource project in opensource community (e.g., GitHub).
The survey is targeted at industry technical experts who have (1) overall architecture design knowledge of the microservice architecture and migration, and (2) experience and familiarity with the main processes of the microservice project. For every survey and interview, a responding practitioner is asked to discuss based on \textbf{one} microservice system that he/she is most familiar with, because a practitioner may be involved in multiple systems that adopt different practices, his/her responses can be vague or conflicting if he/she mixes experience from multiple systems.

The survey questionnaire consists of 26 multiple-choice questions (13 questions with multiple-select checkbox and 2 optional to answer), and 21 Q\&A questions (13 optional to answer). 
The survey includes four main parts:

1) \textbf{Practitioner's Basic Information}:
We collect the basic information of the practitioners, including the name, email address, company, job title, and number of years working in industry.

2) \textbf{Project Overview}:
We collect the project's basic information, including the project area, the number of microservices, LOC, programming languages in use, service origin (legacy vs. new service), and the satisfaction of six benefits listed in Table~\ref{tab:benefit}. 

3) \textbf{Technical Detail Questions}:
We design 34 technical detail questions based on the capabilities described in Section~\ref{sec:concept}. For each capability, we focus on how it has been achieved and the existing shortcomings, or why it is not available.

4) \textbf{Feedback}:
At the end of the questionnaire, we ask for  suggestions and whether they are willing to conduct an interview.


\subsubsection{Interview} 
We received 51 responses of the survey and conducted basic statistical analysis to obtain an overview. 
The data reveals some interesting facts that are at odds with common expectations of microservice systems, e.g., the fact that auto scalability is not achieved in most microservice systems, and that the participants are mostly satisfied with their fault tolerance and fault isolation mechanisms while debugging is recognized as one of the most difficult aspects of microservice systems. 
These responses inspire our curiosity for further inquiry. 
With the understanding that the respondents of the survey may or may not be able to spend enough time to consider each question or provide precise answers carefully, we further invite these participants to conduct face-to-face, in-depth interviews. 
14 of them accepted our invitation.
Through in-depth discussions with these experts from various industry sectors across the globe, we derive our key findings and conclusions.

This stage starts with a brainstorming session based on the survey result, and we decide to conduct a semi-structured interview. 
As recommended by Waring and Wainwright~\cite{EJBRM08ICTUTATCCSFTF}, we prepare a mixture of close-ended and open-ended interview template shown in Figure~\ref{fig:interview-template}. 
The interview template consists of  
(1) information confirmation to warm up and confirm with practitioners the basic information and project overview according to the answers recorded in the questionnaire;  
(2) question setting, i.e., Q1 (in Figure~\ref{fig:interview-template}) aims to confirm the capability within each area; Q2 aims to learn the relevant practices to achieve the capability; Q3 aims to figure out new or unsolved issues and the remaining challenges; and Q4 aims to discover the difficulties (e.g., lack of driving forces or issues) behind capability insufficiency or the reasons why a capability is not required.

To evaluate and improve the interview, we rigorously conducted 2 pilot interviews, resulting in 2 revisions based on the feedback.
Then we conduct these formal interviews mostly (12/14) face-to-face, except two through video conferencing. 
Each interview involves one or two interviewees and three of the authors as the interviewers. 
We record the interview with the consent of the interviewees. During the interview, one interviewer is responsible for asking questions guided by the interview template, while the other interviewers take notes and make sure that all the concerns in this study are covered. 
The interviews are not limited to the template--while trying to learn the adopted practices and issues, we also ask improvisational questions based on the answers of the interviewee(s) to dig deeper~\cite{JAN16SMRDFFQSSI}.  
We conduct 33 follow-up conversations (by phone, email, or in-person) to complement the face-to-face interviews when new issues/practices/challenges are identified later. For example, when we learn that serverless is adopted for rapid response of changes in a later interview, we need to confirm whether the practice is adopted in other systems interviewed before. 
The formal interviews (face to face or online conferencing) and follow-up conversations are all conducted with the same interviewee.
A formal interview may involve several follow-up conversations.

\subsubsection{Analysis} 
With all the data collected and analyzed, we intend to summarize the findings to answer the research questions.

From the survey stage, we use the visual analysis tools provided by the questionnaire platform (e.g., Google Forms) to analyze the answers to each question. 
For closed-ended questions, we directly obtain the required results using the analysis tools.
For open-ended questions, we use the word frequency and perspective analysis tools provided by the platform to assist us in extracting effective information from a large number of texts.
This information helps us identify required categories by open coding and find useful feedback from the participants.
For each open-ended question, three of the researchers code the answers into different categories (e.g., service decomposition methods) through discussion and consensus.

We use the tool of voice-to-text to convert the interview recordings into a text document, and manually proofread the converted document to ensure the accuracy of the data. 
In addition, we summarize the key points of the interview based on the notes and text documents converted from recordings. 
If a recording is not allowed, we still summarize the key points based on the notes. 
Finally, we manually analyze the interview data to answer the three research questions.
For RQ1 (Maturity Level), we extract the promises and benefits (of microservices) that have been fulfilled in the systems and then cluster the systems into different groups to identify the maturity levels.
For RQ2 (Issue), three of the researchers code the mentioned technical topics about various microservice development capabilities into different issues.
For RQ3 (Practice and Challenge), we summarize the practices that have been adopted for each identified issue and estimate the satisfaction levels of the interviewees based on their feedback. 
For an issue that is currently not well addressed, we further consider the challenges behind.

\section{Survey Results}\label{sec:results}

\subsection{Demographics and System Overview}

Based on the 51 responses of the survey we conducted basic statistical analysis to obtain an overview. 
In addition to revealing how technologies are used in different microservice systems, these responses also serve as the basis for us to conduct in-depth conversations with the practitioners. 

\begin{figure}[h]
  \centering
    \begin{tikzpicture}
    \pie[sum=auto, after number=, text=legend, radius=2]{
              4/CTO,
              17/Architect,
              19/development engineer,
              3/DevOps engineer,
              2/Scrum master,
              2/Technical director,
              2/Site Reliability Engineer (SRE),
              1/Project Manager (PM),
              1/Agile coach}
    \end{tikzpicture}
  \caption{Job Tile of Survery}
  \label{fig:jobtitle}
\end{figure}
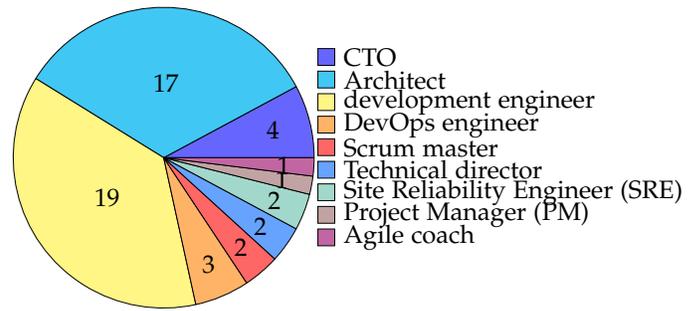
The participants of this survey all play significant roles in developing microservices systems, as shown in Figure~\ref{fig:jobtitle}, including CTO, technical director, architect, scrum master, development engineer, DevOps engineer, and product manager etc. They have 3 to 35 years of professional experience in industry (10.2 years on average).  Each participant fills out a questionnaire based on a microservice system. 

\begin{figure}[h]
  \centering
    \subfigure[The Number of Service]{\label{fig:numOfService}
    \begin{tikzpicture}
         \pie[sum =auto , after number =, radius =1]{3/101-500, 20/ \textless 20 , 26/20-100,  2/ \textgreater 500 }
     \end{tikzpicture}}
     \label{fig:numberOfService}
     \subfigure[Average LOC of Service]{\label{fig:ServiceLOC}
     \begin{tikzpicture}
         \pie[pos ={6 ,0} , sum =auto , after number =, radius=1]{7/ \textless 1000, 20/1001-5000, 11/5001-20000, 13/ \textgreater 20000}
     \end{tikzpicture}}
  \caption{System Scale}
  \label{fig:ServiceSize}
\end{figure}
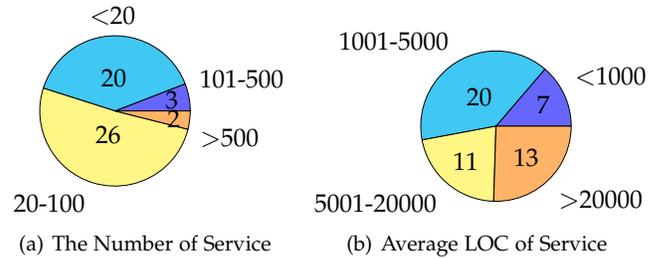

The survey results show that the number of services supported by these microservice systems, ranging from dozens to hundreds. The average number of lines of code (LOC) for each service ranges from hundreds to tens of thousands. We observe that the average number of a service's LOC in a system evolving from a legacy one is larger than that of a newly developed microservice system.


\begin{table*}[h]
  \scriptsize
  \centering
  \caption{The Satisfaction of Microservice Benefits}
  \label{tab:SMBenefits}
  \begin{tabular}{|m{4cm}<{\centering}|m{2cm}<{\centering}|m{1.6cm}<{\centering}|m{1.6cm}<{\centering}|m{1.6cm}<{\centering}|m{1.6cm}<{\centering}|m{1.6cm}<{\centering}|}
  \hline
     \multicolumn{1}{|c|}{\textbf{Benefits}} &  {\textbf{Very Dissatisfied}} & {\textbf{Dissatisfied}}  & {\textbf{Satisfied}}  & {\textbf{Very Satisfied}} & {\textbf{Not care}} & {\textbf{Not know}}\\
     \hline
     Parallel Development  & 0.00\% & 11.76\% & 45.10\% & 43.14\% & 0.00\% & 0.00\% \\
    \hline
     Extendibility and Expandability  & 1.96\% & 11.76\% & 56.86\% & 25.49\% & 0.00\% & 3.92\%  \\
    \hline
     Flexible and Automatic Scalability    & 7.84\% & 35.29\% & 37.25\% & 11.76\% & 5.88\% & 1.96\% \\
    \hline
     Fault Tolerance and Fault Isolation   & 3.92\% & 13.72\% & 58.82\% &  23.53\% & 0.00\% & 0.00\% \\
    \hline
     Reduced Communication Cost  & 1.96\% & 13.72\% & 54.90\% & 23.53\%  & 3.92\% & 1.96\%   \\
    \hline
     Flexible Choice of Technology Stack  & 3.92\% & 15.68\% & 25.49\% & 35.29\% & 13.72\% & 5.88\%  \\
    \hline
  \end{tabular}
\end{table*}

Table~\ref{tab:SMBenefits} summarizes how satisfied the participants are for each of the 6 types of microservice benefits listed in Table~\ref{tab:benefit}. The participants are most (88.24\%) satisfied with \textit{Parallel Development} and least (48.92\%) satisfied with \textit{Flexible and Automatic Scalability}. The data indicate that, in existing microservice systems, parallel development can be achieved very well, but scalability still needs to be improved. It is surprising to observe that 13.72\% participants do not care about the benefit of \textit{Flexible Choice of Technology Stack}.

\subsection{Basic Results}
In this section, we summarize the responses on each technical aspect of microservice systems.

\begin{table}[h]
  \scriptsize
  \centering
  \caption{Service Decomposition Strategies}
  \label{tab:MDStrategies}
  \begin{tabular}{p{4.0cm}|p{0.3cm}<{\centering}}
  \hline
     {\textbf{Strategies}} & {\textbf{\#}} \\
     \hline
     By expert experience & 46 \\
     By Domain-Driven Design & 4 \\
     By data flow & 1 \\
     others & 5 \\
    \hline
  \end{tabular}
\end{table}

\subsubsection{Service Decomposition} 
Service decomposition is the most important part of microservice development, and different systems employ different decomposition strategies, as shown in Table~\ref{tab:MDStrategies}.  
Most (46/51) of the systems decompose services based on expert experience which is an open service decomposition approach and there is no uniform standard and implementation method (Decompose services based on the architect/domain expert's understanding of the business capabilities and domain). Only a few (4/51)systems use a Domain-Driven Design approach to decompose services, including Event Storm and User Journey.
A few participants argue that service decomposition is not designed from the beginning, and needs to evolve according to business changes. Several shortcomings for service decomposition are also mentioned by the participants:
\begin{itemize}
\item Coupling. \textit{"Services are poorly decomposed; there is still some coupling, and some modules interact too frequently and have dependencies on each other..."}
\item Consistency. \textit{"It is difficult to ensure that the service decomposition design is consistent with the code implementation..."}
\item Granularity. \textit{"Service granularity is large; business isolation needs to be improved... Service granularity is too fine; governance costs increase..."}
\end{itemize}

\begin{table}[h]
\scriptsize
  \centering
  \caption{Database Decomposition Strategies}
  \label{tab:DDStrategies}
  \begin{tabular}{p{4.0cm}|p{0.3cm}<{\centering}}
  \hline
     {\textbf{Strategies}} & {\textbf{\#}} \\
     \hline
     By business capability & 27 \\
     By domain & 13 \\
     By horizontal decomposition & 2 \\
     By vertical decomposition & 1 \\
     By data flow & 1 \\
   \hline  
  \end{tabular}
\end{table}  

\subsubsection{Database Decomposition}
It is recommended that in a microservice system, each service should manage its own database~\cite{MICROSERVICE}, but we observe that there are several (7/51) systems using centralized databases, over half (30/51) of the systems have shared databases among services, and only a quarter of (14/51) the systems have no shared database. Table~\ref{tab:DDStrategies} summarizes database decomposition strategies: over half (27/44) of the systems decompose databases based on business capability and a third (13/44) of them decompose databases by domains.

\begin{table}[h]
  \scriptsize
  \centering
  \caption{The Reasons of Shared Database Between Services}
  \label{tab:RSDServices}
  \begin{tabular}{p{7cm}|p{0.3cm}<{\centering}}
  \hline
     {\textbf{Reasons}} & {\textbf{\#}} \\
     \hline
     Business logic is severely coupled & 13\\
     Facilitating data synchronization and cascading queries & 8 \\
     In the process of database decomposing & 3 \\
     Small business scale and low operation and maintenance cost & 1 \\
     Lack of technical support & 1 \\
     Some data source formatting is difficult to solve & 1 \\
    \hline
  \end{tabular}
\end{table}

The reasons why services need to share databases are summarized in Table~\ref{tab:RSDServices}. The main reason is the severe coupling of business logic (13/37), which makes it difficult to decompose databases. Several (8/37) participants reported that a shared database would reduce the time for data synchronization and facilitate cascading queries. In some cases, the databases are decomposed, but not completely, so there are still services sharing one database. Several participants reported shortcomings of database decomposition, with several examples listed below: 
\begin{itemize}
\item Redundancy. \textit{"There is a lot of data redundancy in the data tables..."}
\item Granularity. \textit{"Database granularity is too fine; batch operation performance is low... Database granularity is coarse, not conducive to data isolation..."}
\end{itemize}

\begin{table}[h]
\scriptsize
  \centering
  \caption{The Solutions of Service Deployment}
  \label{tab:Deployment}
  \begin{tabular}{p{5.5cm}|p{0.3cm}<{\centering}}
  \hline
     {\textbf{Solutions}} & {\textbf{\#}} \\
     \hline
     Virtual machine & 14 \\
     Virtual machine, Container & 16 \\
     Container & 12 \\
     Physical machine, Virtual machine & 3 \\
     Physical machine, Virtual machine, Container  & 3 \\
     Physical machine & 2 \\
     Physical machine, Container & 1 \\
   \hline
  \end{tabular}
\end{table}

\subsubsection{Deployment}
A microservice system may contain tens or even hundreds of services. An instance of service can be deployed in a physical machine, a virtual machine, or a container~\cite{DEPLOYMENT}. As reported in Table~\ref{tab:Deployment}, less than half (22/51) of the systems use a hybrid deployment approach: the combination of virtual machines and containers is the most (16/51) popular approach. Some of these systems use only containers (12/51), virtual machines (14/51), or physical machines (2/51). Most systems use Kubernetes (20/51), Mesos (1/51), and Docker Swarm (11/51) to manage containers and use VMware vSphere (15/51) to manage virtual machines. A small number of systems are hosted by cloud platforms (7/51).

\begin{table}[h]
  \scriptsize
  \centering
  \caption{Communication Styles}
  \label{tab:communication}
  \begin{tabular}{p{4.0cm}|p{0.3cm}<{\centering}}
  \hline
     {\textbf{Modes}} & {\textbf{\#}} \\
     \hline
     HTTP/REST, Messaging & 18 \\
     HTTP/REST & 15 \\
     HTTP/REST, RPC, Messaging & 12 \\
     RPC & 4 \\
     HTTP/REST, RPC & 2 \\
    \hline
  \end{tabular}
\end{table}

\subsubsection{Service Communication Design}
In a microservice architecture, services must interact using an inter-process communication protocol such as HTTP, AMQP, and RPC, depending on the nature of each service~\cite{Communication}. There are three main communication styles: HTTP/REST, RPC, and Messaging~\cite{CommunicationStyle}. As shown in Table~\ref{tab:communication}, more than half of (32/51) the systems use a hybrid communication style to meet the requirements of particular scenarios, a third (18/51) of them use both HTTP/REST and Messaging, and a quarter of (12/51) the systems use all of the three communication styles. One-third of the systems employ only one communication style, HTTP/REST (15/51) or RPC (4/51). 


\begin{figure}[h]
  \centering
  \subfigure[Single API Gateway]{
      \begin{minipage}[b]{0.22\textwidth}
        \includegraphics[width=1\textwidth]{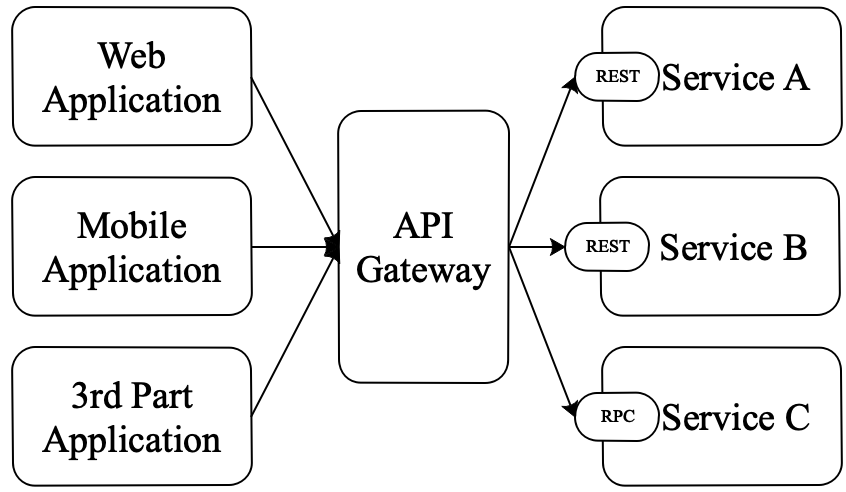}
      \end{minipage}
      \label{fig:SingleAPIGateway}
  }
    \subfigure[Multiple API Gateway]{
      \begin{minipage}[b]{0.22\textwidth}
        \includegraphics[width=1\textwidth]{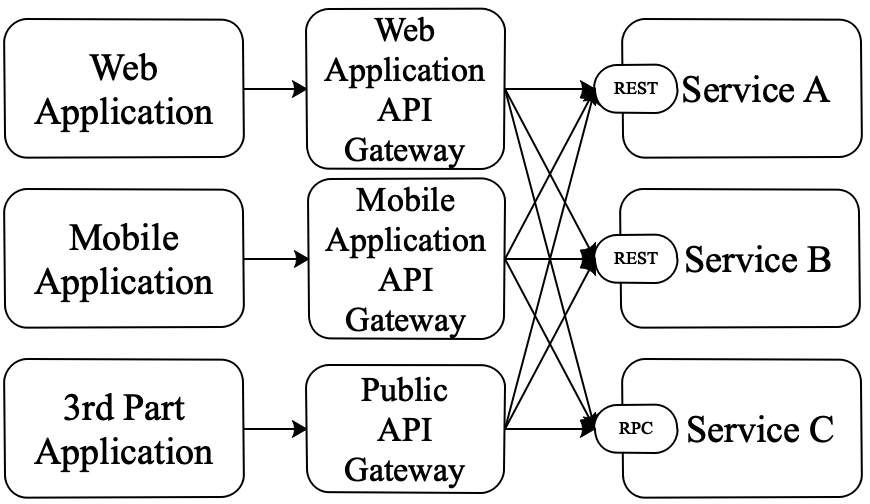}
      \end{minipage}
      \label{fig:MultipleAPIGateway}
    }
  \caption{API Gateway}
  \label{fig:APIGateway}
\end{figure}

\subsubsection{API Gateway Design} %
An API Gateway is a server that is the single entry point into the system. It might have other responsibilities such as authentication, monitoring, load balancing, caching, request shaping and management, and static response handling. There are two kinds of API gateways: single API gateway (Figure~\ref{fig:SingleAPIGateway}) and multiple API gateway  (Figure~\ref{fig:MultipleAPIGateway}, a variation of Backend for Frontend~\cite{bff})~\cite{APIGATEWAY}. The main difference between a single API gateway and a multiple API gateway is that the latter defines different gateways for different clients. In Figure~\ref{fig:MultipleAPIGateway}, there are three kinds of clients: web application, mobile application, and external 3rd party application. There are three different API gateways. Each one provides a set APIs for its clients. 
The survey results show that over half (28/51) of the systems use single API gateway and about a quarter (14/51) of them use multiple API gateway, and only a few (9/51) do not use any API Gateway.

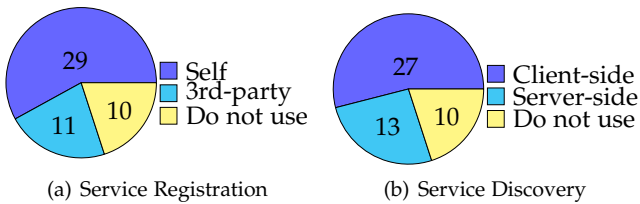
\begin{figure}[h]
  \centering
    \subfigure[Service Registration]{\label{fig:ServiceRegistration}
    \begin{tikzpicture}
         \pie[text = legend, sum =auto , after number =, radius =1]{29/Self, 11/3rd-party, 10/Do not use}
     \end{tikzpicture}}
     \subfigure[Service Discovery]{\label{fig:ServiceDiscovery}
     \begin{tikzpicture}
         \pie[text = legend, sum =auto , after number =, radius=1]{27/Client-side, 13/Server-side, 10/Do not use}
     \end{tikzpicture}}
  \caption{Service Registration and Discovery}
  \label{fig:RegistrationDiscovery}
\end{figure}

\subsubsection{Service Registration and Discovery}
Service registration and discovery are the key components for a microservice system. 
Based on them, client services could then dynamically discover and invoke the required functionalities without any explicit reference to the invoked services' location~\cite{S18JSFCA}.
In a microservice system, service instances have dynamically assigned network locations. Moreover, the set of service instances change dynamically because of autoscaling, failures, and upgrades. The service registration and discovery mechanism helps locate a service instance in a runtime environment. Locating a service instance is divided into two phases: registration and discovery. There are two kinds of registration patterns: self-registration and third-party registration, and also two kinds of discovery patterns: client-side discovery and server-side discovery~\cite{SERVICEDISCOVERY}. According to Figure~\ref{fig:RegistrationDiscovery}, most (40/51) of the systems have a service registration mechanism, including self-registration (29/51) and third-party registration (11/51). A majority (40/51) of the systems use a service discovery mechanism, including client-side discovery (27/51) and server-side discovery (13/51).

\begin{figure*}[t]
    \centering
    \includegraphics[width=12cm]{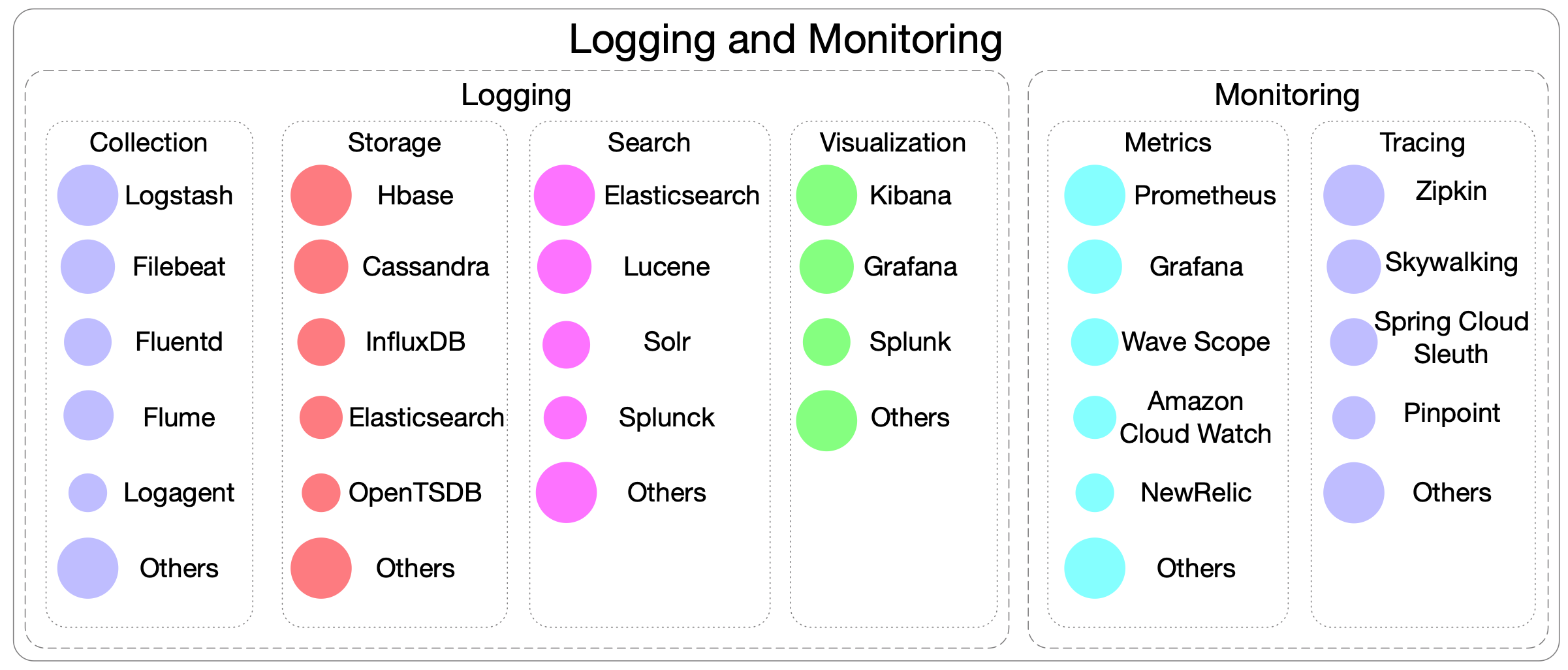}
    \caption{Logging and Monitoring Tools} \label{fig:logmonitor}
\end{figure*}

\subsubsection{Logging and Monitoring}
A microservice system often employs sophisticated monitoring and logging mechanisms to manage individual services using dashboards, so that the  up/down status and a variety of operational and business relevant metrics can be displayed~\cite{MICROSERVICE}. As shown in Figure~\ref{fig:logmonitor}, it is observed that most of the logging and monitoring platforms are built with open-source systems, such as the ELK stack (i.e., Logstash~\cite{logstash} for log collection, ElasticSearch~\cite{elasticsearch} for log indexing and retrieval, Kibana~\cite{kibana} for visualization), and Zipkin~\cite{zipkin} for tracing and visualization.
Some systems use a self-developed Application Performance Management (APM) system or commercial software. Some participants complained about \textit{``insufficient granularity of monitoring metrics and poor support for specific middleware''.}

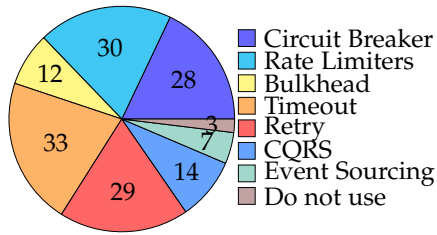
\begin{figure}[h]
  \centering
    \begin{tikzpicture}
    \pie[text = legend ,sum =auto ,  radius=1.5]
        {28/Circuit Breaker, 
        30/ Rate Limiters, 
        12/Bulkhead, 33/Timeout,
        29/Retry, 14/CQRS,
        7/Event Sourcing,
        3/ Do not use}
    \end{tikzpicture}
  \caption{Mechanisms for Performance and Availability Assurance}
  \label{fig:Mechanisms}
\end{figure}

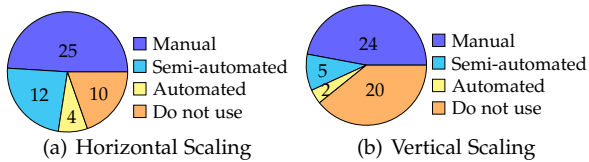
\begin{figure}[h]
  \centering
  \scriptsize
    \subfigure[Horizontal Scaling]{\label{fig:h-scaling}
    \begin{tikzpicture}
        \pie[text = legend ,sum =auto  ,radius=0.8]
        {25/Manual,
        12/ Semi-automated, 
        4/Automated, 
        10/ Do not use}
     \end{tikzpicture}}
     \subfigure[Vertical Scaling]{\label{fig:v-scaling}
      \begin{tikzpicture}
        \pie[text = legend ,sum =auto  ,radius=0.8]{
         24/Manual, 
         5/ Semi-automated, 
         2/Automated, 
         20/ Do not use}
     \end{tikzpicture}}
  \caption{Scaling Strategies}
  \label{fig:Scaling}
\end{figure}

\subsubsection{Performance and Availability Assurance}
There are a large number of remote calls between services, and any request error could cause cascading failures. Remote calls could also lead to additional network latency, affecting system performance. As shown in Figure ~\ref{fig:Mechanisms}, we find that almost all (48/51) of the systems use various strategies to handle the performance and availability issues. Timeout, Rate Limiters, Retry and Circuit Breaker are the most common strategies used in real-world microservice systems, suggesting that more is being done to ensure the availability of network resources and less of the hardware resources. As reported in Figure~\ref{fig:Scaling}, horizontal scaling (41/51) is much preferred to vertical scaling (31/51). In addition, the proportion of automatic scaling (4/51 Horizontal, 2/51 Vertical) is very low, and nearly half (25/51 Horizontal, 24/51 Vertical) are done manually (more details being discussed in Section~\ref{section:performance}).

\begin{table}[h]
\scriptsize
  \centering
  \caption{Testing Approches}
  \label{tab:test}
  \begin{tabular}{p{4.0cm}|p{0.3cm}<{\centering}}
  \hline
     {\textbf{Approaches}} & {\textbf{\#}} \\
     \hline
     Unit testing & 47 \\
     Integration testing & 41 \\
     Stress testing & 37 \\
     End-to-end testing & 35 \\
     Component testing & 24 \\
     Consumer driven contract testing & 13 \\
     Chaos engineering & 6 \\
    \hline
  \end{tabular}
\end{table}

\subsubsection{Testing}%
Due to the dynamics and complexity of microservice systems, testing strategies suitable for monolithic applications need to be reconsidered~\cite{MSTESTING}. Table~\ref{tab:test} shows that unit testing (47/51) is used in almost all systems, followed by integration testing (41/51), stress testing (37/51), end-to-end testing (35/51), component testing (24/51) and consumer-driven contract testing (13/51). We observe that testing whether the business logic meets the requirements is a priority, followed by performance. A few (6/51) of the systems use chaos engineering to ensure the quality of services.

\begin{table}[h]
\scriptsize
  \centering
  \caption{The Approaches of Fault Localization}
  \label{tab:FaultLocalization}
  \begin{tabular}{p{6.0cm}|p{0.5cm}<{\centering}}
  \hline
     {\textbf{Approaches}} & {\textbf{\#}} \\
     \hline
      By logging & 28 \\
      By monitoring tools(metrics,distributed tracing) & 20 \\
      By testing & 5 \\
      By remote debugging  & 5 \\
      By local debugging & 4 \\
      Others & 3 \\
      \hline 
  \end{tabular}
\end{table}  

\subsubsection{Fault Localization}
In the production environment, a large portion of microservice failures are related to the complex and dynamic interactions and dynamic runtime environments~\cite{FSE19LEPFLMA}. Table~\ref{tab:FaultLocalization} shows that over half (28/51) of the practitioners use logs for troubleshooting, two-fifths (20/51) of them use monitoring tools to help locate faults, and the rest (17/51) of the practitioners locate faults by testing, remote/local debugging, etc. Most developers may use a combination of these approaches to locate faults, such as monitoring and distributed tracing. The practitioners also reported some shortcomings in troubleshooting:
\begin{itemize}
\item Disappointing logging and monitoring. \textit{``distributed tracing should be combined with log, process state, and distributed service state to provide richer problem diagnosis information.''}
\item Insufficient automation. \textit{``locating a fault usually requires human effort to look at logs and monitor information on multiple platforms,  lacking effective automation mechanisms to assist in locating faults.''}
\item Fault reproduce. \textit{``production failures are difficult to reproduce in the development environment.''}
\end{itemize}

\begin{table}[h]
\scriptsize
  \centering
  \caption{System Assessment and Measurement}
  \label{tab:assessment}
  \begin{tabular}{p{2.0cm}|p{4.0cm}|p{0.4cm}<{\centering}}
  \hline
     {\textbf{Types}} & {\textbf{Metrics}} & {\textbf{\#}} \\
     \hline
     \multirow{4}{*}{System metrics} &  CPU & 37  \\
      \cline{2-3} & Memory & 33  \\
      \cline{2-3} & Network latency & 28  \\
      \cline{2-3} & I/O & 15 \\
      \cline{2-3} & Thread & 9  \\
      \hline
     \multirow{8}{*}{Service metrics} &  Query Per Second (QPS) & 26  \\
      \cline{2-3} & Transaction Per Service (TPS) & 15 \\
      \cline{2-3} & Error and exception & 8  \\
      \cline{2-3} & Success rate & 7 \\
      \cline{2-3} & Response time & 2  \\
      \cline{2-3} & Call chain statistics  & 4  \\
      \cline{2-3} & Mean Time To Repair (MTTR) & 3  \\
    \hline
  \end{tabular}
\end{table} 

\subsubsection{Service Evolution}
Quality assessment plays an important role in the evolution of microservice systems. As shown in Table~\ref{tab:assessment}, the systems often use two kinds of metrics to assess the quality of the system. Some practitioners mentioned metrics related to operation system and hardware, including CPU, memory, and network latency. A few participants reported that they pay more attention to  service-relevant metrics, such as QPS, TPS, error,  and exception.

\section{Findings}\label{sec:findings}
\begin{table}[t]
\newcommand{\tabincell}[2]{\begin{tabular}{@{}#1@{}}#2\end{tabular}}
  \scriptsize
  \centering
  \caption{Subject Systems in Interviews}
  \label{tab:systems}
    \resizebox{0.47\textwidth}{!} {
   \begin{tabular}{|c|c|c|c|c|c|c|}
  \hline
     {\textbf{Com.}} & {\textbf{Area}} & {\textbf{Sys.}} & {\textbf{Domain}}  & {\textbf{\#Serv}} & {\textbf{KLOC}} & {\textbf{Origin}}  \\
    \hline
        C1 & Finance & S1 & Financial & 20+ & 5+ & New \\
    \hline
        C2 & Internet & S2 & E-Commerce & 10+ & 0.3+  & Legacy \\
    \hline
        C3 & IT & S3 & E-Commerce & 12 & 5+ & New   \\
    \hline
        C4 &  IT & S4 & E-Commerce & 40+  & 0.3+ & New  \\
    \hline
        C5 & IT & S5 & Development & 10+ & 20+ & New\\
    \hline
        C6 & IT & S6 & E-Commerce & 20+ & 100+ & Legacy\\
    \hline
        C7 & Internet & S7 & Entertainment  & 800+ & 50+ & New  \\
    \hline
    \multirow{3}{*}{C8} & \multirow{3}{*}{Internet} & S8 & E-Commerce & 20+ & 100+ & Legacy  \\
      \cline{3-7} && S9 & Development & 10+ & 0.5+ & New   \\
      \cline{3-7} && S10 & Development & 20+ & 2+ & New \\
    \hline
     \multirow{3}{*}{C9} & \multirow{3}{*}{Internet}  & S11 & E-commerce & 100+ & 0.3+ & New  \\
      \cline{3-7} && S12 & Development & 20+ & 10+ & Legacy   \\
      \cline{3-7} && S13 & E-Commerce & 100,000+ & 50+ & Legacy \\
    \hline
     C10 &Manufacture & S14 &Manufacture & 300+ & 50+ & New \\
    \hline
  \end{tabular}
  }
\end{table}

We interviewed the designers of 14 microservice systems from 10 companies of different domains (including finance, Internet, IT, manufacture) as listed in Table~\ref{tab:systems}.
These companies vary in sizes and four of them are Fortune 500 companies.
The systems cover different domains such as finance, e-commerce, entertainment, manufacture, and software development.
Their sizes vary greatly in terms of both service number (10+ to 1000+, S13 is a huge ecosystem that spans multiple business domains, and has 100,000+ services.) and service size (0.3K+ to 100K+ lines of code per service).
Among the 14 systems, 9 are newly developed microservice systems, while the other 5 are evolved from legacy systems.
Based on the survey and interviews, we summarize our findings and answer the three research questions.

\subsection{Maturity Levels (RQ1)}
Our survey and interviews reveal that the promises and benefits of microservices have been more or less fulfilled in these systems.
On the other hand, the degree of fulfillment varies greatly in different systems.
Some basic benefits are well fulfilled in most subject systems, while more advanced benefits are only well fulfilled in a few systems.
Based on the fulfillment degree of the benefits defined in Table~\ref{tab:benefit}, we group the subject systems into the following three levels.
Note that a company may have multiple microservice systems (e.g., S11-S13) of different levels.

\begin{itemize}
\item \textbf{Level 1 - Independent Development and Deployment}.
The development and deployment of different services are isolated, and each service can be developed and deployed independently in an autonomy way.

\item \textbf{Level 2 - High Scalability and Availability}.
The system can be flexibly scaled with system load and other environmental changes, and at the same time ensure high availability through fault tolerance and fault isolation.

\item \textbf{Level 3 - Service Ecosystem}.
The system has been evolved into an ecosystem that can support not only the extension of new requirements but also the expansion of business domains.

\end{itemize}

S1-S4 belongs to level 1.
These systems usually have dozens of services.
They implement some basic characteristics of microservices: each service is physically isolated, running in its own process and communicating with lightweight mechanisms~\cite{MICROSERVICE, S18JSFCA}.
All the systems are deployed on physical or virtual machines.
S2 also uses containers for some services but without container orchestration.
To deploy or manage the systems, the operators usually need to allocate resources manually (e.g., virtual machines), configure runtime environments, and upgrade or downgrade packages.
These practices are called mutable infrastructure~\cite{IMMUTABLE}, which means that the infrastructure will be continually updated, tweaked, and tuned to meet the ongoing needs of the purpose it serves.
Mutable infrastructure is known to suffer from a number of problems.
For example, it is hard to scale, as each service instance's creation involves a lot of manual configurations~\cite{ImmutableInfrastructure};
it is hard to recover or rollback from failures, as the configuration of a service instance is unknown after changes~\cite{ImmutableInfrastructure}.
All these systems implement continuous integration/deployment (CI/CD) pipelines except S4.

To evolve a monolithic system to a level 1 microservice system, the organization needs to refactor the system into a set of services, form a cross-functional team for each service, and establish CI/CD pipelines.
It is usually challenging to conduct this kind of large-scale architectural refactoring~\cite{FSE16IGARFSR} and at the same time ensure that the system behaviors are not changed.

S5-S12 belong to level 2.
These systems usually have dozens to hundreds of services.
All of these systems are deployed with lightweight containers (e.g., Docker) and some of them use a hybrid deployment of virtual machines and containers.
A common characteristic is that they have established the practices of immutable infrastructure~\cite{IMMUTABLE}, where service instances are replaced rather than changed.
The practices are based on the construction and deployment of images (e.g., Docker images).
Moreover, all these systems implement DevOps practices except S5, which just implements CI/CD pipelines.

To evolve a level 1 system to a level 2 system, the organization needs to establish a series of infrastructures, e.g., container and image management and even container orchestrators (e.g., Kubernetes), and runtime monitoring systems.

S13 and S14 belong to level 3.
These systems have been evolved into service ecosystems consisting of hundreds to thousands of services.
In these systems, services from different business domains are interconnected and supported by a set of common infrastructures.
The service infrastructures offer not only technical supports (e.g., resource allocation and scaling, database management, and message queue) but also business supports such as the accesses of business services (e.g., user authorization and authentication).
These systems adopt more advanced cloud computing technologies. For example, service mesh~\cite{ServiceMesh} provides the capability of traffic management and decouples the business logic and infrastructure, allowing developers to focus on their business logic without being distracted by business-neutral issues, such as Service Discovery, Circuit Breaker, and Rate Limiters. Serverless~\cite{SERVERLESS} includes BaaS (Backend as a Service)~\cite{BAAS} and FaaS (Function as a Service)~\cite{FAAS}.
BaaS provides a set of common services including technical services (such as user authentication and authorization service) and business services (such as payment service), and we can quickly build new services and even new applications based on BaaS.
FaaS is about realizing customized business logic by running backend code without managing your own server systems or your own long-lived server applications, speeding up the new-idea-to-initial-deployment story.

To evolve a level 2 system to a level 3 system, the organization needs to improve the technical infrastructures that are expected to provide common business services, introduce new technologies to the infrastructures (e.g., service mesh and Serverless), and establish governance mechanisms for the service ecosystem.

\begin{table}[t]
  \scriptsize
  \centering
  \caption{Benefits Fulfilled by Different Maturity Levels}
  \label{tab:benefitstar}
  \begin{tabular}{|m{4.3cm}<{\centering}|m{0.9cm}<{\centering}|m{0.9cm}<{\centering}|m{0.9cm}<{\centering}|}
    \hline
    \textbf{Benefit}  & \textbf{Level 1} & \textbf{Level 2} & \textbf{Level 3} \\
    \hline
    Parallel Development & \score{1.5}{3} & \score{2}{3} & \score{2.5}{3}\\
    \hline
    Extendibility and Expandability & \score{1.5}{3} & \score{1.5}{3} & \score{2.5}{3}\\
    \hline
    Flexible and Automatic Scalability & \score{1}{3} & \score{2}{3} & \score{2.5}{3}\\
    \hline
    Fault Tolerance and Fault Isolation & \score{2}{3} & \score{2.5}{3} & \score{2.5}{3}\\
    \hline
    Reduced Communication Cost & \score{2}{3} & \score{2}{3} & \score{2}{3}\\
    \hline
    Flexible Choice of Technology Stack & - & - & -\\
    \hline
  \end{tabular}
\end{table}

The three maturity levels fulfill different benefits of microservices as shown in Table~\ref{tab:benefitstar}.
We score points at different maturity levels for each benefit.
It is a subjective assignment based on the following criteria: 
a delta of a half star and one star denote minor and fundamental improvement over a lower level, respectively.

\textbf{Parallel Development}.
The benefit is primarily related to service granularity and largely fulfilled at level 1 with the decomposition and isolation of services. It can be seen from Table~\ref{tab:SMBenefits} that the practitioners are most (88.24\%) satisfied with \textit{Parallel Development}.
With the advances in infrastructures, level 2 and level 3 systems can better support the development and frequent updating of smaller services (e.g., functions in FaaS), and thus can better fulfill the benefit.

\textbf{Extendibility and Expandability}.
At level 1 and level 2, the benefit is fulfilled based on the proper decomposition of services, easing the extension of new services or requirements.
For example, the application of DDD (Domain Driven Design) makes services better aligned with domain concepts and thus facilitate requirements extension~\cite{IEEESoft18CDMDAMP}.
Level 3 systems further facilitate the emergence of new applications based on the business supports of service infrastructures, thus support the expansion of business domains.

\textbf{Flexible and Automatic Scalability}.
At level 1, the operators manually scale the services based on their experiences.
Level 2 systems implement semi-automatic scaling of services:
monitoring mechanisms alarm the operators on possible performance degradation and the operators make service scaling decisions based on the alarms. The auto-scaling mechanisms provided by the container orchestrators or cloud platforms automatically execute the scaling instructions.
Level 3 systems fulfill better scalability based on Serverless, which enables the deployment and delivery of fine-grained service functionalities without creating and managing the required infrastructure resources~\cite{S18JSFCA}.
It is worth noting that no systems in our study implement fully automatic scaling of services due to the concern about the controllability and risks~\cite{DDOSSCALING}.

\textbf{Fault Tolerance and Fault Isolation}.
Level 1 systems largely fulfill fault tolerance and fault isolation based on the physical isolation of services (ensuring that the failures of any service do not affect other services) and fault tolerance mechanisms such as timeout, retry and circuit breaker~\cite{FaultTolerence}.
Level 2 and level 3 systems better fulfill fault tolerance and fault isolation based on failure recover and rollback mechanisms supported by immutable infrastructures. 
In addition, the systems can more easily realize blue green deployment~\cite{BLUEGREEN}, canary release (including A/B testing)~\cite{CANARY} and rolling update~\cite{ROLLINGUPDATE}, so that the systems can be migrated to the new version fast and smoothly.

\textbf{Reduced Communication Cost}.
The benefit is primarily related to physical isolation of services, explicit definition of service contracts, and cross-functional teams.
It is largely fulfilled at level 1 and has no significant changes at level 2 and 3.

\textbf{Flexible Choice of Technology Stack}.
The aspect mainly relies on the scale of the organization and is irrelevant to the maturity levels.
Hybrid technology stacks, e.g., using multiple languages in service development, are expensive to maintain for small organizations.
These technology stacks require the organization to have developers familiar with different technology stacks and a common dependency being implemented in different languages. 
This requirement explains why 13.72\% participants in the survey do not care about the benefit of \textit{Flexible Choice of Technology Stack} (see Table~\ref{tab:SMBenefits}).
Our interviews show that large organizations tend to choose hybrid technology stacks.

The fulfillment of the benefits is driven by specific business forces, and the driving force at a higher level includes the driving forces of lower levels.
The driving force of level 1 is the fast response to market changes and business innovations,  necessitating independent development and deployment of services.
The driving force of level 2 is highly available services for a large number of user accesses,  necessitating flexible and automatic scalability.
The driving force of level 3 is business expansion and merger, necessitating service ecosystems and high expandability.


\subsection{Issues, Practices, and Challenges (RQ2 \& RQ3)}
\begin{table*}[h]
  \footnotesize
  \centering
  \caption{Sources of Issues, Practices, and Challenges (I, P, C mean the issue, practice, challenge, respectively)}
  \label{tab:IssueSource}
  \resizebox{\textwidth}{!}{
  \begin{tabular}{|c|c|ccc|ccc|ccc|ccc|ccc|ccc|ccc|ccc|ccc|ccc|ccc|}
  \hline
  \multicolumn{2}{|c|}{\multirow{2}{*}{\textbf{System}}} 
      & \multicolumn{3}{c|}{\textbf{Issue 1}}
      & \multicolumn{3}{c|}{\textbf{Issue 2}}
      & \multicolumn{3}{c|}{\textbf{Issue 3}}
      & \multicolumn{3}{c|}{\textbf{Issue 4}}
      & \multicolumn{3}{c|}{\textbf{Issue 5}}
      & \multicolumn{3}{c|}{\textbf{Issue 6}}
      & \multicolumn{3}{c|}{\textbf{Issue 7}}
      & \multicolumn{3}{c|}{\textbf{Issue 8}}
      & \multicolumn{3}{c|}{\textbf{Issue 9}}
      & \multicolumn{3}{c|}{\textbf{Issue 10}}
      & \multicolumn{3}{c|}{\textbf{Issue 11}}\\
  \cline{3-35}
      \multicolumn{2}{|c|}{}&{\textbf{I}} & {\textbf{P}} & {\textbf{C}}
      & {\textbf{I}} & {\textbf{P}} & {\textbf{C}}
      & {\textbf{I}} & {\textbf{P}} & {\textbf{C}}
      & {\textbf{I}} & {\textbf{P}} & {\textbf{C}}
      & {\textbf{I}} & {\textbf{P}} & {\textbf{C}}
      & {\textbf{I}} & {\textbf{P}} & {\textbf{C}}
      & {\textbf{I}} & {\textbf{P}} & {\textbf{C}}
      & {\textbf{I}} & {\textbf{P}} & {\textbf{C}}
      & {\textbf{I}} & {\textbf{P}} & {\textbf{C}}
      & {\textbf{I}} & {\textbf{P}} & {\textbf{C}}
      & {\textbf{I}} & {\textbf{P}} & {\textbf{C}}\\
  \hline
  \multicolumn{2}{|c|}{Survey}
      &$\surd$&$\surd$&$\surd$
      & & & 
      &$\surd$ & & 
      &$\surd$&$\surd$&
      &$\surd$& & 
      & & & 
      & & & 
      & & &
      & & & 
      &$\surd$&$\surd$& 
      & & &\\
  \hline
  \multirow{4}{*}{Level 1} & S1 
      &$\surd$& & 
      &$\surd$&$\surd$&$\surd$ 
      & & & 
      &$\surd$&$\surd$&$\surd$
      &$\surd$& & 
      & & & 
      & & & 
      &$\surd$&$\surd$&$\surd$
      &$\surd$&$\surd$&$\surd$ 
      &$\surd$&$\surd$&$\surd$
      & & &\\
  \cline{2-35} & S2
      &$\surd$& & 
      & & & 
      &$\surd$&$\surd$&$\surd$ 
      &$\surd$&$\surd$&$\surd$ 
      & & & 
      &$\surd$&$\surd$& 
      & & & 
      & & & 
      & & & 
      & & & 
      & & &\\
  \cline{2-35} & S3
      &$\surd$& &
      & & & 
      & & & 
      &$\surd$&$\surd$&$\surd$ 
      & & & 
      & & & 
      & & & 
      & & & 
      &$\surd$&$\surd$&$\surd$ 
      & & & 
      & & &\\
  \cline{2-35} & S4
      &$\surd$&$\surd$&$\surd$ 
      &$\surd$& & 
      & & & 
      & & &
      &$\surd$& & 
      & & & 
      & & & 
      &$\surd$& & 
      &$\surd$&$\surd$&$\surd$ 
      &$\surd$&$\surd$&$\surd$  
      & & & \\
  \hline
  \multirow{8}{*}{Level 2} & S5
      &$\surd$&$\surd$&$\surd$ 
      & & & 
      & & & 
      &$\surd$&$\surd$&$\surd$
      &$\surd$& & 
      & & & 
      & & & 
      & & & 
      & & & 
      & & & 
      & & & \\ 
  \cline{2-35} & S6 
      &$\surd$& & 
      &$\surd$&$\surd$&$\surd$ 
      &$\surd$&$\surd$&$\surd$ 
      &$\surd$&$\surd$&$\surd$ 
      & & & 
      &$\surd$&$\surd$&$\surd$ 
      & & & 
      &$\surd$&$\surd$&$\surd$ 
      &$\surd$&$\surd$&$\surd$ 
      &$\surd$&$\surd$&$\surd$ 
      & & & \\
  \cline{2-35} & S7 
      & & & 
      &$\surd$&$\surd$&$\surd$ 
      & & & 
      &$\surd$&$\surd$&$\surd$ 
      &$\surd$& & 
      & & & 
      & & & 
      &$\surd$&$\surd$&$\surd$ 
      &$\surd$&$\surd$&$\surd$ 
      &$\surd$&$\surd$&$\surd$ 
      &$\surd$&$\surd$&$\surd$ \\
  \cline{2-35} & S8 
      & & & 
      &$\surd$&$\surd$&$\surd$ 
      & & & 
      &$\surd$&$\surd$&$\surd$ 
      & & & 
      & & & 
      &$\surd$&$\surd$&$\surd$ 
      &$\surd$&$\surd$&$\surd$
      &$\surd$&$\surd$&$\surd$
      &$\surd$&$\surd$&$\surd$ 
      & & &\\
  \cline{2-35} & S9 
      &$\surd$&$\surd$&$\surd$ 
      & & & 
      & & & 
      & & & 
      & & & 
      & & & 
      &$\surd$&$\surd$&$\surd$ 
      & & &
      &$\surd$&$\surd$&$\surd$
      &$\surd$&$\surd$&$\surd$ 
      & & & \\
  \cline{2-35} & S10 
      &$\surd$&$\surd$&$\surd$ 
      &$\surd$&$\surd$&$\surd$ 
      & & & 
      & & & 
      & & & 
      & & & 
      &$\surd$&$\surd$&$\surd$ 
      &$\surd$&$\surd$&$\surd$
      &$\surd$&$\surd$&$\surd$
      &$\surd$&$\surd$&$\surd$ 
      & & & \\
  \cline{2-35} & S11 
      & & & 
      &$\surd$&$\surd$&$\surd$ 
      & & & 
      &$\surd$&$\surd$&$\surd$ 
      & & & 
      & & & 
      &$\surd$&$\surd$&$\surd$ 
      &$\surd$&$\surd$&$\surd$
      &$\surd$&$\surd$&$\surd$
      &$\surd$&$\surd$&$\surd$
      & & & \\
  \cline{2-35} & S12 
      & & & 
      & & & 
      &$\surd$&$\surd$&$\surd$  
      & & & 
      &$\surd$& & 
      & & & 
      &$\surd$&$\surd$&$\surd$ 
      & & & 
      &$\surd$&$\surd$&$\surd$ 
      &$\surd$&$\surd$&$\surd$ 
      & & & \\
  \hline
  \multirow{2}{*}{Level 3} & S13 
      &$\surd$&$\surd$&$\surd$ 
      &$\surd$&$\surd$&$\surd$
      &$\surd$&$\surd$&$\surd$
      &$\surd$&$\surd$&$\surd$
      &$\surd$& & 
      &$\surd$&$\surd$&$\surd$
      &$\surd$&$\surd$&$\surd$
      &$\surd$&$\surd$&$\surd$
      &$\surd$&$\surd$&$\surd$
      &$\surd$&$\surd$&$\surd$
      & & & \\
  \cline{2-35} & S14 
      & & & 
      &$\surd$&$\surd$&$\surd$
      & & & 
      &$\surd$&$\surd$&$\surd$
      & & & 
      & & &
      &$\surd$&$\surd$&$\surd$
      &$\surd$&$\surd$&$\surd$
      &$\surd$&$\surd$&$\surd$
      &$\surd$&$\surd$&$\surd$
      &$\surd$&$\surd$&$\surd$  \\
  \hline
  \end{tabular}}
\end{table*}

For each of the 11 capabilities, we investigated possible restrictive issues, practices that have been adopted to address the issues, and the remaining challenges.

The investigation revealed 11 issues, 10 practices, and 10 challenges, which emerge from the survey and interviews (Table~\ref{tab:IssueSource}). 
5 issues (i.e., Issues 1, 3, 4, 5, 10) are mentioned both in the survey and in the interviewed systems; the other 6 only showed up in the interviewed systems.
The higher the maturity level, the more issues a system has to face, and the more practices need to be adopted.
 
There are 4 capabilities that have no issues: communication design, API gateway design, service registration and discovery, and testing.
This result indicates that the organizations generally have no difficulties in these aspects with the support of mature tools and infrastructures.
For example, service registration and discovery are well supported by Eureka~\cite{Eureka} and Zookeeper~\cite{Zookeeper}.
We next report the issues identified for each capability together with the practices addressing each issue and the remaining challenges.

\subsubsection{Service Decomposition}
The capability of service decomposition is restricted by the issues on decomposition decision, design evaluation, and refactoring of legacy systems.

\textbf{1) Issue 1: Decomposition Decision Influences a Lot}.
Services in a microservice system are physically isolated as basic units of development, deployment, and scaling.
Improper service decomposition may cause serious quality problems (e.g., performance and scalability).
On the other hand, physical isolation makes it impossible to access the internal logics of a service from outside. 
Therefore, improper service decomposition may hinder the implementation of new requirements.
Some interviewees reported that they had to merge multiple services together due to these problems.
Moreover, they reported that the refactoring of microservice systems is much harder than that of monolithic systems, as the refactoring crosses the boundaries of multiple services with independent development teams.
This issue influences all the three maturity levels in a similar way.

\textbf{Practice: Domain Driven Design (DDD)}. 
DDD~\cite{DDDdef} is a software development methodology that focuses on mapping concepts in the problem domain into artifacts in the solution domain.
Although DDD is not specific for microservices, it is widely used in microservices systems to achieve more effective service decomposition.
Actually all the interviewees mentioned that they are using or want to use DDD, but only a few of the systems successfully applied DDD.  

\textbf{Challenge: Domain Model and Artifact Mapping}.
The challenges with DDD lie in the derivation of the domain model and its mapping with artifacts, including 
(1) how to conduct domain analysis to derive a proper domain model;
(2) how to extract domain concepts from the artifacts (e.g., code, test cases) of legacy systems and link them to corresponding new artifacts; 
(3) how to monitor and maintain the consistency between domain model and artifacts.


\textbf{2) Issue 2: Service Dependencies are Hard to Capture}.
Compared with the modularity of monolithic systems, service decomposition quality is more difficult to evaluate.
The modularity evaluation of monolithic systems usually relies on static dependencies and evolutionary coupling of files~\cite{ICSE16IQAD, ICSE16DLNMAMC}.
To use static analysis to capture service dependencies for microservice systems, one needs to map the service IP addresses or service names in the code to service repositories.
This mapping is usually complex and error-prone.
On the other hand, as services are independently developed in separate repositories, it is infeasible to capture evolutionary coupling by analyzing revision histories. 
The missing of service dependencies makes it hard to measure service coupling and further evaluate the quality of service decomposition.
This issue influences all the three maturity levels similarly.

\textbf{Practice: Capturing Service Dependency using Runtime Tracing}.
Runtime tracing captures service invocations and thus can be used to capture service dependencies by analyzing runtime invocations.

\textbf{Challenge: High Cost and Low Coverage of Runtime Tracing.}
Runtime tracing heavily relies on monitoring infrastructure (see Section~\ref{sec:monitor}).
If an organization has not established required infrastructure, it is expensive to implement runtime tracing in an ad hoc way (e.g., by instrumenting monitoring code in an intrusive way).
On the other hand, service dependencies captured by runtime tracing are usually incomplete due to low coverage of potential dependencies.


\textbf{3) Issue 3: Legacy Systems are Hard to Migrate}.
5 out of the 14 interviewed systems are evolved from legacy systems (see Table~\ref{tab:systems}).
Due to the complex dependencies between files, it is often hard to migrate a monolithic legacy system to microservice architecture and at the same time ensure the continuous provision of business services.
This issue mainly influences level 1, as incremental migration to microservice systems usually occurs at this level.

\textbf{Practice: Strangler Pattern and Anti-Corruption Layer Pattern}.
These two patterns are often used together to migrate a legacy system to microservice architecture incrementally.
The Strangler pattern~\cite{STRANGLER} suggests gradual replacement of specific pieces of functionalities with newly developed services. 
The migration process can take a long time, during which both the legacy system and the new services are used to support the business together.
As all the legacy system are replaced by new services, it is eventually strangled and replaced by a microservice system and thus can be decommissioned.
The Anti-corruption Layer~\cite{ACL} pattern suggests isolating the new services from the legacy system by placing an anti-corruption layer between them. 
This layer translates communications between the two parts, allowing the legacy system to remain unchanged while the new services can avoid compromising its design decisions.

\textbf{Challenge: Legacy System Boundary and Halfway Migration}.
It is often hard to incrementally determine a proper boundary between the legacy system and new services.
An improper boundary may make the new services hard to develop (e.g., due to complex dependencies with the legacy system) or the anti-corruption layer hard to implement.
Another challenge lies in halfway migration: the migration process often ends with a big monolithic legacy subsystem after all the easy parts have been replaced with services.
The remaining legacy subsystem is hard to migrate and at the same time causes new problems of dead code, as it may include a large amount of code that has been reimplemented in new services.

\subsubsection{Database Decomposition} 
The capability of database decomposition is restricted by the issue of data coupling among services.

\textbf{Issue 4: Data Coupling among Services}. 
Multiple services may have data coupling when the data elements (e.g., fields or tables) in their databases are involved in the same data query or transaction.
Due to the physical isolation, cascading query and traditional transaction management cannot be used for microservice systems. 
This limitation explains why 37 out of the 51 systems in the survey share databases among services, including 7 using centralized databases and 30 sharing databases among some services.
Note that by sharing databases we mean sharing database servers among different services.
This issue influences all the three maturity levels.

\textbf{Practice: Service Invocation Composition and Distributed Transaction}. 
A cascading query can be implemented by the composition of multiple service invocations.
Distributed transactions and data consistency can be ensured by using distributed transaction frameworks (e.g., Seata~\cite{Seata}).

\textbf{Challenge: Subsequent Refactoring and Network Latency}. 
When a database is decomposed into multiple parts,  the source code that relies on multiple parts of the database (e.g., due to cascading queries) needs to refactored accordingly.  
On the other hand, the composition of multiple service invocations may cause serious network latency.


\subsubsection{Deployment} 
The capability of deployment is restricted by the issue of complex service configurations

\textbf{Issue 5: Complex Service Configurations}. 
Microservice systems usually involve complex service configurations.
For example, improper or inconsistent service configurations (e.g., inconsistent memory limitations of JVM and Docker) often cause runtime failures~\cite{TSE18MS}.
This issue mainly influences level 2 and 3 systems.
Their runtime environments are highly dynamic and it is hard to form a common recommendation for service configurations.
Moreover, it is harder to locate and fix the failures caused by configuration problems.

\textbf{Practice and Challenge}. 
There are no effective practices being identified in this study.
It remains a great challenge to determine and maintain proper configurations for services.


\subsubsection{Performance and Availability Assurance} \label{section:performance}
The capability of performance and availability assurance is restricted by the issues of stateful service and autoscaling strategy.

\textbf{1) Issue 6: Inconsistency across Stateful Services}.
The scaling of a stateful service may cause its multiple instances in inconsistent states, which may in turn cause failures.
Stateful services are not recommended, but  still exist in microservice systems due to the incomplete migration from monolithic to microservice architecture or the choice of inexperienced developers.
This issue mainly influences level 2 and 3 systems.

\textbf{Practice: Migrating States to External Storage}.
By migrating service states to external storage such as in-memory cache (e.g., Redis~\cite{redis}), we can change a stateful service into a stateless one.

\textbf{Challenge: High Refactoring Cost, Network Latency, and System Bottleneck}.
Eliminating service states requires additional refactoring efforts.
On the other hand, external storage may lead to network latency and become system bottleneck due to shared access.

\textbf{2) Issue 7: Unpredictable and Uncontrollable Autoscaling Strategy}.
Autoscaling strategies are hard to test and the effects of the strategies are highly unpredictable.
Moreover, the effects may be uncontrollable.
For example, when a microservice system encounters DoS attacks, the rapid growth of service instances may occupy a lot of resources and thus make services unavailable. 
This limitation explains why the participants in the survey are the least satisfied (7.84\% very dissatisfied, 35.29\% dissatisfied) with \textit{Flexible and Automatic Scalability} (see Table~\ref{tab:SMBenefits}).
This issue mainly influences level 2 and 3 systems.

\textbf{Practice: Semi-automated Scaling}. 
Semi-automated scaling alarms the operators on possible problems, and the operators make scaling decisions.
The decisions are then automatically executed, e.g., by creating more service instances.

\textbf{Challenge: Predictable and Reliable Autoscaling.}
Although none of the interviewed systems implement fully automatic scaling, it is still desired.
Predictable and reliable autoscaling requires intelligent scaling decision making and sound quality assurance at runtime.


\subsubsection{Logging and Monitoring}\label{sec:monitor}
The capability of logging and monitoring is restricted by the issues on distributed tracing and service anomaly detection.

\textbf{1) Issue 8: Complex and Asynchronous Service Invocation Chain}. 
Microservice systems often involve complex and asynchronous service invocation chains.
Distributed tracing is usually required to pinpoint where failures occur and what causes poor performance~\cite{TRACING}.
Its impact increases from level 1 to level 3.
Level 2 systems usually have more complex and dynamic service invocation chains.
Level 3 systems crossing multiple business domains have more complex service interactions.

\textbf{Practice: Invasive and Non-invasive Tracing}. 
There are two types of approaches for distributed tracing:
invasive approaches implement tracing by instrumenting probes into services; 
non-invasive approaches implement tracing by using sidecar to proxy network requests in a service mesh.

\textbf{Challenge: High Cost, Fragility, and Latency}. 
The challenges with invasive approaches include 
(1) the cost of code instrumentation is high, especially when there are a lot of services written by hybrid languages;
(2) the tracing chains are fragile, as any problems (e.g., missing or wrongly passed trace ID) with the probe of a service may cause the whole tracing chain to be interrupted or corrupted.
The challenges with non-invasive approaches include 
(1) the cost for the required infrastructures such as service mesh;
(2) network latency caused by sidecar proxies.

\textbf{Issue 9: Service Incidents are Hard to Detect.}
Due to the dynamic behaviors (e.g., scaling) and high granularity of microservices, detecting service incidents is more challenging than before. This issue exists across all maturity levels. Level 2 and level 3 systems are more dynamic and flexible, so they get impacted more than level 1.

\textbf{Practice: Dashboard and Threshold}. 
Many systems conduct anomaly detection by manually analyzing the metrics provided by the dashboard of monitoring systems (e.g., Grafana and Prometheus), or setting thresholds to trigger predefined actions.

\textbf{Challenge: Insufficient Automation}.
Operators cannot be on the dashboard 24 hours a day; therefore, anomaly detection tools with full or high automation are needed.
The dashboard cannot display all the metrics of interest, and operators often need to cross multiple monitoring platforms to obtain the required information. 
In addition, it is difficult to ensure the accuracy and timeliness of the thresholds, because threshold setting depends on experience and it needs to be constantly adjusted for system changes.


\subsubsection{Fault Localization}
The capability of fault localization is restricted by the issue of complex and dynamic service interactions.

\textbf{Issue 10: Complex and Dynamic Service Interaction.}
Faults and failures of a microservice system often involve complex and dynamic distributed environments and service invocation chains.
For example, service instances are dynamically created and destroyed, and the analysis of a request execution process often requires logs distributed in many service instances.
This issue often makes it hard to identify a failing circumstance and reproduce it in the test environment.
The issue exists across all maturity levels, but influences more on level 2 and 3 systems.
These systems are more dynamic and involve more complex runtime environments and service invocation chains, so they get impacted more than level 1.

\textbf{Practice: Local Debugging, Mock, Remote Debugging, Traffic Routing}. 
There are different practices applicable to different levels.
Level 1 systems can use local debugging to locate faults, as the invocation chains usually involve only a small number of services. 
For level 2 and level 3 systems, local debugging is usually infeasible, as the debugging may involve many services and the local environment cannot host all of them.
These systems can use remote debugging by connecting to a remote server and using online debugging tools (e.g., Visual Studio Remote Debugger~\cite{VSRemoteDebugger}).
An alternative practice is traffic routing~\cite{TrafficManagement}, which routes service traffic from a remote environment (e.g., a test environment) to the local environment.

\textbf{Challenge: Debugging Performance, Infrastructure Requirements, and Lack of Intelligence}. 
Remote debugging may not be smooth due to the communications with a remote server, which may become a bottleneck.
Traffic routing requires the supporting network infrastructures (e.g., Sidecar~\cite{Sidecar}), which are not available for many systems.
Overall, the current microservice fault localization practices lack intelligence and automation (e.g., machine-learning-based fault localization approaches~\cite{FSE19LEPFLMA}) are highly desired.


\subsubsection{Service Evolution} 
The capability of service evolution is restricted by the issue on evolution compatibility.

\textbf{Issue 11: Evolution Compatibility}.
A microservice system may have a lot of services and the upgrade of a service may cause compatibility problems with upstream services.
Its impact increases from level 1 to level 3.
Level 2 systems are more dynamic and more frequently upgraded based on the support of immutable infrastructures.
Service evolution in level 3 systems has much broader ranges of change impact in ecosystems.

\textbf{Practice: Downward Compatibility and Upgrade Deadline}.
A common practice for this issue is to ensure downward compatibility by providing multiple versions of a service API. 
This multi-version API provision usually can be done by adding a version prefix to the URI (Uniform Resource Identifier).
At the same time, the usages of different service API versions are continuously monitored and those that are no longer used can be decommissioned.
The organizations may also set an upgrade deadline; after the deadline passes, service API versions will no longer be available.

\textbf{Challenge: High Maintenance Cost}. 
The impact of a service upgrade is often hard to predict, so the developers are not sure whether other services will be broken by the upgrade.
Therefore, the developers may choose to keep many different versions of the same service API in the production environment, thus causing high maintenance costs.

\section{Discussion}\label{sec:discuss}
The complexity and dynamism of microservice systems pose unique challenges to a variety of software engineering tasks~\cite{TSE18MS}.
Learning the roadmap of industrial microservice systems and grounding the practices and challenges on the maturity levels can help researchers better understand potential research opportunities in the context of different maturity levels.

The focus of level-1 microservice systems is independent development and deployment, and the aim of this stage is to establish a design structure that conforms to the microservice architecture.
For newly developed systems, the challenges are mainly related to service decomposition and service coupling.

Software design methodologies such as domain-driven design (DDD) and distributed transaction are widely practiced.
However, improper service decomposition and data coupling still often prevent the system from \realizing the desired independence, extensibility, and performance.
DDD is a successful methodology, but lacks effective techniques and tools.
Knowledge-based techniques and tools are required to map between the domain model and artifacts, and maintain their consistency, e.g., by extracting domain concepts from the artifacts~\cite{fse19chong}.
On the other hand, microservice architecture analysis techniques need to consider the independence and extensibility, e.g., by assessing the alignment of the domain model and architecture model,
and also the performance, e.g., by estimating the frequency and latency of distributed service invocations.

For migrated microservice systems, the challenges are mainly related to the incremental migration process.
Although the strangler pattern and anti-corruption layer pattern are used to support incremental migration, it is often the case that a big monolithic legacy subsystem remains in the system and interacts with migrated services.
The reason why the developers choose to keep the monolithic subsystem is usually that it is not cost effective to refactor the subsystem.
Microservice refactoring techniques are required to not only provide migration suggestions but also safely implement the refactoring. 

Moreover, this hybrid architecture brings additional difficulties to runtime monitoring and tracing.
The monolithic subsystem influences the observability of the whole system, as the subsystem encapsulates a large unobservable part of the system and its old technology stack makes it hard to apply the latest tracing frameworks.

The focus of level-2 microservice systems is high scalability and availability, and the aim of this stage is to establish the operation infrastructure and practices required by high-scalability and high-availability microservice systems.

The challenges are mainly related to the detection and localization of faults.
Some organizations have advocated AIOps (artificial intelligence for IT operations), and intelligent fault detection and localization are  important parts of its practices.
Microservice systems at this level widely adopt distributed tracing, and thus fault detection and localization can utilize not only logs and metrics but also traces.
The analysis involves a huge amount of data.
For example, a large Internet system may produce billions of traces per day.
Highly efficient data analysis techniques are thus required to detect and locate potential faults among a large number of services, using techniques such as data mining, machine learning, and interactive visualization.
Moreover, analyzing the combination of logs, metrics, and traces is challenging, as they are produced at different levels: 
logs reflect the local behaviors of individual service instances; 
metrics measure the availability of infrastructure resources and quality of services;
traces record the invocation chains of services for requests.
An integrated representation of the data is required to support effective and efficient fault detection and localization.

The focus of level-3 microservice systems is service ecosystem, and the aim of this stage is to establish the technical infrastructures that support the continuous expansion of business domains.
The challenges are mainly related to the required domain abstraction and service governance mechanisms. 
Like software product lines, service ecosystems are constructed based on a set of core assets that embody the commonality of the domains.
Different from traditional software product lines, service ecosystems rely on microservice infrastructure and common services to implement the core assets. 
A prominent challenge is how to elicit and construct a stable abstraction for an open and uncertain future.
The challenge with service governance originates from the large number of services and the complex service interactions across multiple relevant domains.
The applications in a service ecosystem may emerge continuously based on existing infrastructure and common services.
To ensure the sustainable evolution of the ecosystem, it is thus crucial to establish a comprehensive set of mechanisms to support service governance requirements such as configuration management, authorization, versioning, and evolution management.

\section{Threats to Validity}\label{sec:threat}

In our study, there are three main threats to internal validity. 
The first one is the reasonability of the question settings: some questions are optional or designed to solicit in-depth answers, so our key insights are discovered and validated during the interviews. 
The second one lies in the qualification of the interviewees: we list preferred qualifications in the survey and select interviews based on their answers. One interview may invite more than one interviewee, and follow-up conversations are conducted to ensure the quality of interviews. 
The third one lies in the validity of the answers from interviewees: for the reputation of their company, some interviewees may not fully faithfully provide answers from their real experience.

There are two main threats to the external validity of our study. 
The first one is the quantitative limitation of the surveys and interviews. Our study targets at the experts who must have sufficient knowledge and experience in microservice practices.
Microservice experts with full-stack knowledge and experiences are rare.
51 participants in the survey is not an ideal response rate, but still forms a meaningful sample space. 
In addition, we carefully selected the interviewee to avoid noise data and maintain a diversified group for the generalizability. The second one lies in the limited variety of interviewees' role. We target at architects and experienced developers mainly because they have the most comprehensive background. There is a lack of data from operation and management personnel.

\section{Related Work}\label{sec:related}
Recently, there have been various investigations on the practice of microservice architecture.
Hassan \emph{et al.}~\cite{SCC16MTDTOSAR} formulated the problem of addressing the microservice design trade-offs and introduced their solution proposal.
Phipathananunth \emph{et al.}~\cite{COMPSAC18SRMMSA} described Pink, a framework for synthetic runtime monitoring of microservices software systems. 
Pina \emph{et al.}~\cite{ISNCA18NMMS} proposed a much simpler and non-invasive monitoring approach that includes topology and performance metrics.
Zhou \emph{et al.}~\cite{ASE18DDMS} presented a debugging approach for microservice systems based on the delta debugging algorithm, which is to minimize failure inducing deltas of circumstances for effective debugging.
Du \emph{et al.}~\cite{IC3APP18ADDCMPM} designed an anomaly detection system (ADS) to detect and diagnose anomalies in microservices by monitoring and analyzing real-time performance data.
Zhou \emph{et al.}~\cite{FSE19LEPFLMA} proposed MEPFL, an approach of latent error prediction and fault localization for microservice systems by learning from system trace logs.  
Each of the preceding efforts mainly focused on a specific proposed practice of microservices. In this work, we focus on learning industrial challenges, practices, and capabilities from real-world examples of microservice systems.

Dragoni \emph{et al.}~\cite{CoRR16MYTT} reviewed the development history from objects, services, to microservices, presented the current state-of-the-art and raised some open problems and future challenges. 
Francesco \emph{et al.}~\cite{ICSA17RAMTFPIA} performed a systematic mapping study to identify, classify, and evaluate the current state-of-the-art on architecting microservices from the following three perspectives: publication trends, focus of research, and potential for industrial adoption.
Pahl \emph{et al.}~\cite{CCSS16MSMS} conducted a systematic mapping study on the motivation, architecture, methods, techniques, and challenges of microservices.
Alshuqayran \emph{et al.}~\cite{SOCA16SMSMA} conducted a systematic study on identifying architectural challenges, the architectural diagrams/views, and quality attributes related to microsevice systems. 
Jamshidi \emph{et al.}~\cite{S18JSFCA} reported the current situation, benefits, evolution, and future challenges of microservices. 
Carlos \emph{et al.}~\cite{ECASE17BRMSR} presented an initial set of requirements for a candidate microservice benchmark system to be used in research on software architecture. 

These previous research efforts conducted valuable systematic studies from the literature, but do not represent the practices in industry. Our work conducts an extensive questionnaire survey along with interviews with industrial microservice experts, and can help better understand the state-of-the-practice and the real challenges remaining to be addressed for future research.

Taibi \emph{et al.}~\cite{ICC17PMIMMAAEI} conducted a questionnaire survey filled by 21 practitioners and focused on processes, motivation, and issues for migration toward a microservice architecture. 
Francesco \emph{et al.}~\cite{ICSA18MTMAAIS} performed an empirical study on migration practices toward the adoption of microservices in industry. They collected information utilizing interviews and questionnaires on the activities and the challenges during the migration.
Stefan \emph{et al.}~\cite{SOCA18EISAMD} investigated the importance of different areas of microservice design. Ten microservice experts were interviewed to understand the importance and relevance of the microservices design areas.
Justus \emph{et al.}~\cite{ICSA-C19MIITCSQ} contributed a qualitative study with insights into industry adoption and implementation of microservices by analyzing 14 service-based systems during 17 interviews. They focused on applied technologies, microservices characteristics, and the perceived influence on software quality.
Zhang \emph{et al.}~\cite{ICSA19MARII} carried out a series of industrial interviews with 13 different types of companies to investigate the gap between the ideal visions and real industrial practices and the benefits of microservices from the industrial experiences. 
Ford~\cite{TSMM} conducted a survey on the state of microservices practices in industry, including continuous deployment and automated testing, containers and kubernetes, integration with legacy applications.
Taibi \emph{et al.}~\cite{Taibi2020} identified a taxonomy of 20 anti-patterns, including both organizational ones and technical ones.
Rechards \emph{et al.}~\cite{msa-antipattern-pitfalls} reported 10 common microservice anti-patterns and pitfalls.
Taibe \emph{et al.}~\cite{IEEESoftware18ODMBS} collected evidence of microservice-specific bad practices and classified them into a catalog of 11 microservice-specific bad smells.

These preceding research efforts reported insightful findings on microservice systems in practice but do not classify or investigate the immanent patterns through multi-dimensional diversities. Our work not only derives the maturity levels of microservice systems but also draws a comprehensive roadmap associated with inherent challenges. Our work also explores potential research opportunities by observing industry experiences and obtains a lot of in-depth findings on practice patterns and ecosystem characteristics.


\section{Conclusion}\label{sec:conclude}
In this article, we have reported an empirical study, including an online survey and a series of interviews, that is designed to advance our understanding of microservice practices in industry and remaining challenges that may lead to valuable research. 
As a part of our findings, we have identified three maturity levels of microservice systems: independent development and deployment, high scalability and availability, and service ecosystem, based on the \realized benefits of microservices. 
We have also identified 11 practical issues that restrict the microservice capabilities of organizations and the corresponding practices and challenges.

For practitioners, our findings can help them position their microservice systems at a proper level according to the business needs and determine what infrastructures and capabilities are worth investing. 
More concretely, they can learn the issues that restrict their capabilities and the practices that they can follow to address the issues. 
In this way, they can set a roadmap of continuous improvement.
For researchers, our findings can help them understand the situation of industrial microservice practices and the needs of microservice systems of different levels. 
They may identify potential research opportunities to address the challenges in the current microservice practices.

In future work, we plan to investigate the research problems identified in this work and at the same time work together with our industrial collaborators to establish a shared benchmark for microservice research.
In addition, we plan to further explore the emerging technical problems in service ecosystems.





\ifCLASSOPTIONcaptionsoff
  \newpage
\fi

\bibliographystyle{IEEEtran}
\bibliography{IEEEabrv,reference}


\end{document}